%% file: main.tex
\documentclass[sigconf]{acmart}

\AtBeginDocument{%
  \providecommand\BibTeX{{%
    \normalfont B\kern-0.5em{\scshape i\kern-0.25em b}\kern-0.8em\TeX}}}

\setcopyright{none}

\renewcommand\footnotetextcopyrightpermission[1]{} 

\usepackage{booktabs}   
\usepackage{subcaption} 

\usepackage{xcolor}
\usepackage{soul}
\usepackage{xspace}
\usepackage[epsilon]{backnaur}
\usepackage{framed}
\usepackage{listings}
\usepackage{balance}
\usepackage{amsmath,amsfonts,amsbsy,bm,stmaryrd}
\usepackage[mathscr]{eucal}
\usepackage{xparse}
\usepackage{algorithmic}
\usepackage{graphicx}
\usepackage{wrapfig}
\usepackage{algorithm} 
\usepackage{wrapfig}
\usepackage{xargs}
\usepackage{subcaption}
\usepackage{cleveref}
\usepackage[acronym]{glossaries}
\makeglossaries

\usepackage{nicematrix}

\usepackage{tikz}
\usetikzlibrary{matrix}
\usepackage{multirow}
\usepackage{tabularx}
\makeatletter
\def\hlinewd#1{%
\noalign{\ifnum0=`}\fi\color{white}\hrule\@height #1 %
\futurelet\reserved@a\@xhline}
\makeatother

\newlength{\Oldarrayrulewidth}

\def\paperversiondraft{draft}

\NewDocumentCommand \T { O{} m } {\ensuremath{\boldsymbol{#1\mathscr{\MakeUppercase{#2}}}}}
\NewDocumentCommand \M { O{} m } {\ensuremath{\bm{#1\mathbf{\MakeUppercase{#2}}}}} 
\NewDocumentCommand \V { O{} m } {\ensuremath{\bm{#1\mathbf{\MakeLowercase{#2}}}}} 

\definecolor{ruiqin}{RGB}{23,125,54}
\definecolor{gokcen}{RGB}{255,50,255}
\definecolor{lenny}{RGB}{255,165,0}
\definecolor{jiajia}{RGB}{255, 111, 97}
\definecolor{bin}{RGB}{50,125,154}

\ifx\paperversiondraft\undefined
\newcommand{\rt}[2][1=]{}
\newcommand{\gk}[2][1=]{}
\newcommand{\lgg}[2][1=]{}
\newcommand{\jl}[2][1=]{}
\newcommand{\br}[2][1=]{}
\else
\newcommandx{\rt}[2][1=]{#1\textcolor{ruiqin}{\textbf{Ruiqin:} #2}}
\newcommandx{\gk}[2][1=]{#1\textcolor{gokcen}{\textbf{Gokcen:} #2}}
\newcommandx{\lgg}[2][1=]{#1\textcolor{lenny}{\textbf{Lenny:} #2}}
\newcommandx{\jl}[2][1=]{#1\textcolor{jiajia}{\textbf{Jiajia:} #2}}
\newcommandx{\br}[2][1=]{#1\textcolor{bin}{\textbf{Bin:} #2}}
\fi

\newcommand{\name}{COMET\xspace}%

\usepackage{mips}
\usepackage{color}

\definecolor{dkgreen}{rgb}{0,0.6,0}
\definecolor{gray}{rgb}{0.5,0.5,0.5}
\definecolor{mauve}{rgb}{0.58,0,0.82}

\definecolor{mygreen}{rgb}{0,0.6,0}
\input{acronyms}
\makeglossaries

\usepackage[most]{tcolorbox}
\usepackage{colortbl}
\usepackage{hhline}
\usepackage{array}

\usepackage{tikz}
\usepackage{pgfplots}
\usetikzlibrary{positioning,arrows}
\usetikzlibrary{decorations.pathreplacing,calc}
\newcommand{\tikzmark}[1]{\tikz[overlay,remember picture] \node (#1) {};}
\newcommand*{\addnote}[4]{%
    \begin{tikzpicture}[overlay, remember picture]
        \draw [decoration={brace,amplitude=0.5em},decorate,thick,brown]
            ($(#3)!([yshift=1.5ex]#1)!($(#3)-(0,1)$)$) --
            ($(#3)!(#2)!($(#3)-(0,1)$)$)
                node [align=center, text width=1.2cm, pos=0.5, anchor=west] {#4};
    \end{tikzpicture}
}%

\begin{document}


\title{A High-Performance Sparse Tensor Algebra Compiler in Multi-Level IR}

\author{Ruiqin Tian}
\affiliation{%
 \institution{Pacific Northwest National Laboratory}
 \country{}}
\email{ruiqin.tian@pnnl.gov}

\author{Luanzheng Guo}
\affiliation{%
 \institution{Pacific Northwest National Laboratory}
 \country{}}
\email{lenny.guo@pnnl.gov}

\author{Jiajia Li}
\affiliation{%
 \institution{Pacific Northwest National Laboratory}
 \country}
\email{Jiajia.Li@pnnl.gov}

\author{Bin Ren}
\affiliation{%
 \institution{William \& Mary}
 \country{}}
\email{bren@cs.wm.edu}

\author{Gokcen Kestor}
\affiliation{%
 \institution{Pacific Northwest National Laboratory, UC Merced}
 \country{}}
\email{gokcen.kestor@pnnl.gov}








\input{abstract}

\pagestyle{plain}

\maketitle

\input{intro}

\input{background}
\input{overview}
\input{tstorage}
\input{language}

\input{codegen}

\input{opt}
\input{evaluation}

\input{related_work}

\input{conclusion}

\section{Acknowledgement}
This research is supported by PNNL Laboratory Directed Research and Development Program (LDRD), Data-Model Convergence Initiative, project DuoMO: A Compiler Infrastructure for Data-Model Convergence.

\bibliographystyle{ACM-Reference-Format}
\bibliography{refs}



\printglossary 

\end{document}

%% file: acronyms.tex
\newacronym{IR}{IR}{Intermediate Representation}
\newacronym{CPU}{CPU}{Central Processing Unit}
\newacronym{LLVM}{LLVM}{Low-Level Virtual Machine}
\newacronym{MLIR}{MLIR}{Multi-Level Intermediate Representation}
\newacronym{DSL}{DSL}{Domain-Specific Language}
\newacronym{CSR}{CSR}{Compressed Sparse Row}
\newacronym{DCSR}{DCSR}{Double Compressed Sparse Row}
\newacronym{CSF}{CSF}{Compressed Sparse Fiber}
\newacronym{COO}{COO}{COOrdinate}
\newacronym{SpMM}{SpMM}{Sparse-Matrix Dense-Matrix}
\newacronym{SpMV}{SpMV}{Sparse-Matrix Dense-Vector}
\newacronym{GPU}{GPU}{Graphics Processing Unit}
\newacronym{TA}{TA}{Tensor Algebra}
\newacronym{AI}{AI}{Artificial Intelligence}
\newacronym{SCF}{SCF}{Structured Control Flow}

%% file: abstract.tex
\begin{abstract}
Tensor algebra is widely used in many applications, such as scientific computing, machine learning, and data analytics. 
The tensors represented real-world data are usually large and sparse.
There are tens of storage formats designed for sparse matrices and/or tensors and the performance of sparse tensor operations depends on a particular architecture and/or selected sparse format, which makes it challenging to implement and optimize every tensor operation of interest and transfer the code from one architecture to another.
We propose a tensor algebra domain-specific language (DSL) and compiler infrastructure to automatically generate kernels for mixed sparse-dense tensor algebra operations, named \name{}.
The proposed DSL provides high-level programming abstractions that resemble the familiar Einstein notation to represent tensor algebra operations. The compiler performs code optimizations and transformations for efficient code generation while covering a wide range of tensor storage formats.
\name{} compiler also leverages data reordering to improve spatial or temporal locality for better performance.
Our results show that the performance of automatically generated kernels outperforms the state-of-the-art sparse tensor algebra compiler, with up to 20.92x, 6.39x, and 13.9x performance improvement, for parallel SpMV, SpMM, and TTM over TACO, respectively.

\end{abstract}

%% file: intro.tex
\section{Introduction}
Tensor algebra is at the core of numerous applications in scientific computing, machine learning, and data analytics.
Tensors are a generalization of matrices to any number of dimensions, which are often large and sparse.
Sparse tensors are used to represent a multifactor or multirelational dataset, and has found numerous applications in data analysis and mining~\cite{kolda2008scalable,rendle2009learning,song2019tensor} for health care~\cite{acar2007multiway,luo2017tensor}, natural language processing~\cite{bouchard2015matrix,otter2020survey}, machine learning~\cite{li2017mlog,sidiropoulos2017tensor}, and social network analytics~\cite{zhang2014cap}, among many others.

Developing optimized kernels for sparse tensor algebra methods is complicated. First, sparse tensors are often stored in a compressed form (indexed data structures) and computational kernels needs to efficiently loop over the nonzero elements of the tensor inputs.
Second, iterating over nonzero elements highly depends on the particular storage format employed, hence many algorithms exist to implement the same operation, each targeting a specific format.
Finally, applications may use multiple formats concurrently throughout the computation and mix different formats in the same operation to achieve high performance. 
When tensors with different storage formats are used in the same operation, there are two options: converting one (or both) tensor(s), which is time-consuming especially if the tensor is only used once, or developing an algorithm that can efficiently iterate over both formats simultaneously, which lacks generality and requires different implementations for each combination of tensor formats~\cite{baskaran2012efficient,liu2017unified}.

The current solutions implement ad hoc high-performance approaches for particular computer architecture and/or format.
Most of these algorithms tackle specific problems and domains and conveniently store sparse tensors in a format that exploits the characteristics of the problem. 
This approach has originated tens of different formats~\cite{smith2015splatt,li2018hicoo,nisa2019efficient,choi2010model,baskaran2012efficient} to represent sparse tensors.
Some of them are storage-efficient for specific inputs~\cite{choi2010model,vuduc2005fast,kourtis2011csx,baskaran2012efficient} or evenly nonzero distributions across rows/columns~\cite{monakov2010automatically,choi2010model}; some are better affiliated to specific tensor computations, e.g., sparse matrix-vector multiplication~\cite{yan2014yaspmv,xie2018cvr} versus sparse tensor-matrix multiplication~\cite{smith2017accelerating,baskaran2012efficient}; others are particularly designed for different computer architectures, such as CPUs~\cite{xie2018cvr,li2013smat} versus GPUs~\cite{maggioni2013adell,merrill2016merge}. 
On the other hand, it is infeasible to manually write optimized code for each tensor algebra expressions considering the all possible combinatorial combinations of tensor operations and formats.

To solve the above challenges, we present a sparse tensor algebra compiler, named \name, that is agnostic to storage formats: as opposed to a library of sparse tensor methods, where the methods are statically defined, a compiler can automatically and dynamically generate efficient tensor algebra kernel specifically optimized mixed dense/sparse tensor expressions.
\name{} \gls{DSL} is a highly-productive language that provides high-level programming abstractions that resemble the familiar Einstein notations~\cite{einstein1923grundlage} to represent tensor operations. 
\name{} is based on the \gls{MLIR}~\cite{lattner2020mlir} framework recently  introduced  by  Google  to building reusable and extensible compiler infrastructures.  
The key benefit of building on top of MLIR is its built-in performance portability.
In  the  \name{}  multi-level \gls{IR}, domain-specific,  application-dependent  optimizations  are  performed  at  higher levels  of  the \gls{IR}  stack  where  operations  resemble  programming  languages’  abstractions  and  can  be  optimized  based  on  the  operations  semantics. Generic, architecture-specific optimizations are, instead, performed at lower-levels, where simpler operations are mapped to the memory hierarchy and to processor’s registers.


To enable modular code generation with respect to formats and combination of formats, 
we employ four storage format attributes -- \emph{dense, compressed unique, compressed non-unique, and singleton} -- which are assigned to each tensor dimension~\cite{kjolstad:2017:taco}. By properly combining those attributes in each dimension, it is possible to express common sparse tensor compressed formats, such as COO, CSR, DCSR, ELLPACK, CSF and Mode-generic. 
\name{} code generation algorithm analyzes the dimension attributes and produces code to efficiently iterate over the nonzero elements of the input tensors. 
Since the number of storage format attributes is far lower than all possible combinations of storage formats, the code generation algorithm is greatly simplified and yet can support most of the commonly used sparse tensor storage formats and arbitrary combinations of those. 
This approach lets users not only mix and match storage format desired for their applications but also can enable custom formats without modifying the underlying compiler infrastructure.
Once the loop form of a computation has been generated at the \gls{IR}, \name{} either lowers the code for sequential or parallel execution. In the former case, \name{} produces a high-quality LLVM \gls{IR} (which we show in this work has better loop unrolling and vectorization than an equivalent LLVM \gls{IR} produced by \texttt{clang}); in the latter case, instead, \name{} lowers code to the async dialect for asynchronous task execution based on LLVM co-routines 
Compared to hand-tuned libraries~\cite{li2013smat,bell2009implementing,yan2014yaspmv,li2018hicoo,smith2015splatt,nisa2019efficient} and source-to-source compilers~\cite{kjolstad:2017:taco,kjolstad:2017:tacotool,kjolstad2019tensor}, our approach is more portable, flexible, and adaptable, as emerging architectures and storage formats can be added without re-engineering the computational algorithms.
Finally, \name{} employs the state-of-the-art data reordering algorithm~\cite{li2019efficient} to increase spatial and temporal locality on a modern processor.

We evaluated \name{} with $2833$ sparse matrices and six tensors from the SuiteSparse Suite Matrix Collection~\cite{davis2011university}, FROSTT Tensor Collection~\cite{frosttdataset} and BIGtensor~\cite{haten2_ICDE2015}.
Our results show that \name{} can generate efficient code for multi-threaded CPU architectures from high-level descriptions of the algorithms. 
Compared to state-of-the-art high-productivity tensor algebra languages and compiler, \name{} provides on average 2.29x, up to 6.26x, performance improvements over TACO compiler for sequential \gls{SpMM}.
We also show that asynchronous task execution outperforms OpenMP parallelization, especially for small input matrices, where runtime overhead is predominant. Our results show up to 6.39x and 13.9x speedup over TACO for SpMM and TTM, respectively.
Finally, data reordering achieves up to 3.89x and 7.14x performance improvements for parallel SpMV and SpMM kernels, respectively, over the original \name{}.


To the best of our knowledge, \name{} is the \emph{first} \gls{MLIR}-based compiler that integrates generic code generation for arbitrary input formats, data reordering, and automatic parallelization within the same framework. \name{} can improve end-user application performance while supporting efficient code generation for a wider range of formats specialized for different application and data characteristics.
This paper makes the following contributions: 
\begin{itemize}
    \item We introduce the \name{} \gls{DSL}, an intuitive yet powerful and flexible language to implement dense and sparse tensor algebra algorithms; 
    \item We propose an \gls{MLIR}-based compiler that automatically generates efficient sequential and parallel code for a tensor expression with dense and mixed operands while supporting the important sparse tensor storage formats.
    \item We integrate the state-of-the-art data reordering algorithm to enhance data locality. 
    \item We provide an exhaustive experimental evaluation and show that \name{} generally outperforms state-of-the-art tensor compiler for both sequential and parallel execution. 
\end{itemize}

%% file: background.tex
\section{Background and Motivation}
\label{sec:background}
There exist various compressed and uncompressed formats to store sparse matrices and tensors in the literature, including \gls{COO}, \gls{CSR}, \gls{DCSR}, ELLPACK, \gls{CSF}, and Mode-Generic~\cite{bell2009implementing,feng2011optimization,merrill2016merge,chen2020aesptv, kincaid1989itpackv}.  
The specific format chosen to represent data in an application generally depends on the expected characteristics of the data itself and how these impact other desired properties, such as performance of a computational kernel or memory footprint (which is particularly important in the case of very large, multi-dimensional tensors).

Each format is important for different reasons. 
\gls{COO}~\cite{sedaghati2015automatic,anzt2020load} is commonly used to store sparse matrices and tensors, such as the Matrix Market exchange format~\cite{matrixmarket} and the FROSTT sparse tensor format~\cite{frosttdataset}.
While COO is the most natural format, it is not necessarily the most performant format.
\gls{CSR}~\cite{white1997improving} is for sparse matrices, which compresses row indices as pointers to row beginning positions to avoid duplicated storage and increase performance for memory bandwidth-bound computation such as \gls{SpMV}.
\gls{DCSR}~\cite{buluc2008representation} further compresses zero rows by adding an extra pointer to nonzero rows based on the CSR format. With an extra level of compression on rows, \gls{DCSR} is more efficient than \gls{CSR} for highly sparse (hypersparse) data.
The ELLPACK~\cite{kincaid1989itpackv} format is efficient for matrices that contain a bounded number of nonzeros per row, such as matrices that represent well-formed meshes.
\gls{CSF}~\cite{smith2015splatt} generalizes the \gls{DCSR} or \gls{CSR} matrix format to high-order tensors that compresses every dimension.
Mode-Generic format~\cite{baskaran2012efficient} is a generic representation of semi-sparse tensors with one or more dense dimensions stored as dense blocks with the coordinates of the blocks stored in \gls{COO}.

An application might need any or even several of these formats based on its needs, which makes it important to support computation with various tensor storage formats and  their combinatorial combinations. 
The main challenge is that the computational kernel needs to effectively iterate over each sparse input tensor stored in different storage formats.
This problem is especially more complicated for expressions that involve multiple operands. 

Because of the large number of storage formats and possible combinations, most state-of-the-art sparse tensor libraries support only a few sparse formats (and generally only binary operations) or convert tensors to an internal storage format, thereby potentially losing the performance, memory footprint, or other advantages that a specific format may offer.
A compiler, on the other hand, can automatically generate the efficient code for specific input formats and their combinations, increasing flexibility, adaptivity to new formats, and portability to various hardware platforms. To achieve this goal, two important requirements need to be satisfied: 1) a unified way to represent important sparse storage formats (Section~\ref{sec:format}) and 2) an efficient algorithm to generate specific code for a given expression and its particular input formats (Section~\ref{sec:codegen}).


%% file: overview.tex
\begin{figure}
  \centering
  \begin{lstlisting}[xleftmargin=.02\textwidth,numbers=left,deletekeywords={b}, abovecaptionskip=0pt, caption={An example \texttt{SPACe} program for Sparse Matrix-times-Dense-Matrix operation.}, captionpos=b, label={fig:example_program}]
def main() {
  #IndexLabel Definition
  IndexLabel [a] = [?];
  IndexLabel [b] = [?];
  IndexLabel [c] = [32]; 

  #Tensor Definition
  Tensor<double> A([a,b],CSR);  #Tensor<double> A([a,b],{D,CU}); 
  Tensor<double> B([b,c],Dense);#Tensor<double> B([b,c],{D,D});
  Tensor<double> C([a,c],Dense);#Tensor<double> C([a,c],{D,D});

  #Tensor Readfile Operation
  A[a,b] = space_read(filename); 

  #Tensor Fill Operation
  B[b,c] = 1.0;
  C[a,c] = 0.0;

  #Tensor Contraction
  C[a, c] = A[a,b] * B[b,c];
}
  \end{lstlisting}
\end{figure}

\section{\name{} Overview}
\label{sec:overview}


\name{} consists of a DSL for tensor algebra computations, a progressive lowering process to map high-level operations to low-level architectural resources, a series of optimizations performed in the lowering process, and various IR dialects to represent key concepts, operations, and types at each level of the multi-level IR.
This section reviews the key characteristics of our compiler framework. \name{} is based on the \gls{MLIR} framework~\cite{lattner2020mlir}, a compiler infrastructure to build reusable and extensible compilers and IRs. MLIR supports the compilation of high-level abstractions and domain-specific constructs and provides a disciplined, extensible compiler pipeline with gradual and partial lowering. 
Users can build domain-specific compilers and customized IRs (called \emph{dialect}), as well as combining existing IRs, opting into optimizations and analysis. 
 
Our previous work focuses on \emph{dense} high-dimensional tensor contractions. The compiler reformulates tensor contractions 
as a sequence of transpose and matrix-matrix multiplication operations, then generates efficient code by several code optimizations (e.g., loop tiling, micro kernel). 
The detailed description of previous work and its performance results for important tensor expressions from the Northwest Chemistry framework (NWChem)~\cite{valiev2010nwchem} 
can be found in~\cite{blind}.
This work, instead, focuses on \emph{sparse} tensor algebra. 

Figure~\ref{fig:example_program} shows an example \name program for an SpMM operation. 
The \texttt{IndexLabel} operation defines an index label. It can assign the size of the index with a scalar number. If the size is unknown in static time, then use a question mark (?) (Lines 3-5). The \texttt{Tensor} operation defines new tensors
(Lines 8-10); the SpMM operation is defined at Line 20. 
In particular, the matrix \texttt{A} is stored in the CSR format while the matrix \texttt{B} and the result matrix \texttt{C} are dense.  
Note that there is no specific operation for SpMM at language level, nor the programmer needs to explicitly state the format of each input tensor while contracting the two tensors. 
\name atomically derives the specific operation from the format of input tensors and the index labels.
\name{} internally annotates each tensor with storage format attributes, devises the storage formats used in the contraction, and properly passes this information down to the IR stack when lowering the code. \name{} can generate the appropriate code according to the input tensor storage formats (Section~\ref{sec:codegen_alg}).

\begin{figure}[t]
  \begin{center}
    \includegraphics[width=0.45\textwidth]{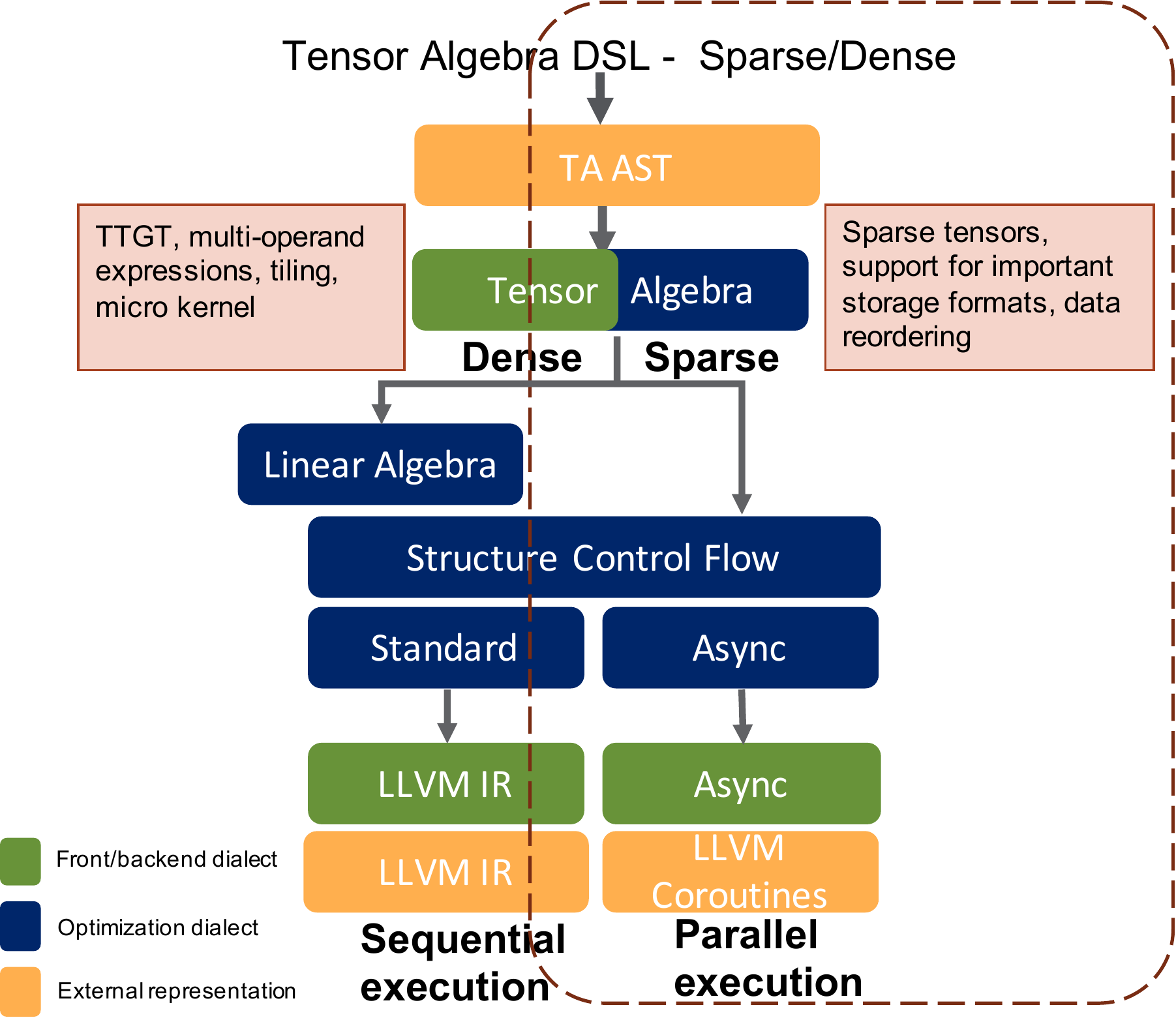}
  \end{center}
  \caption{\name{} execution flow and compilation pipeline}
  \label{fig:overview}
\end{figure}

The code generation in \name{} follows a progressive lowering approach where optimizations are applied at different levels. Figure~\ref{fig:overview} shows the compilation pipeline of \name{}, where our contributions are annotated by the dashed box.
Users express their computation in a high-level tensor algebra DSL (Section~\ref{sec:language}).
First, the \name{} DSL is lowered to a Sparse Tensor Algebra (TA) \gls{IR}, the first dialect in the \name{} IR stack. The language operators, types, and structures are first mapped to an abstract syntax tree and then to the TA dialect. The TA dialect contains domain-specific concepts, such as multi-dimensional tensors, contractions, and tensor expressions. 
Our compiler framework applies high-level optimizations and transformation leveraging semantics information carried from the DSL. For example, \name{} tracks the input tensors' definitions and annotates each tensor with storage format attributes on each dimension, based on the index label definitions. 

Next, our compiler lowers the \gls{TA} \gls{IR} code to lower levels of the \gls{IR} stack, which follows different paths depending on the operation and input formats. Dense tensor algebra operations are lowered first to linear algebra dialect, then to \gls{SCF} dialect, and finally to standard dialect. 
Sparse linear algebra operations are lowered to \gls{SCF} dialect which is a loop represented in the \gls{MLIR} framework.
At this point, \name{} employs generic optimizations during the lowering steps but also considers additional information about the final target architecture. For CPU execution, the code is lowered to the \gls{LLVM} dialect for sequential execution and async dialect to models asynchronous execution at a higher-level and then to proper \gls{LLVM} \gls{IR} for final assembly and linking. 

%% file: tstorage.tex
\begin{figure*}[h]
    \subfloat[Sparse Matrix]{
        \includegraphics[width=0.48\textwidth]{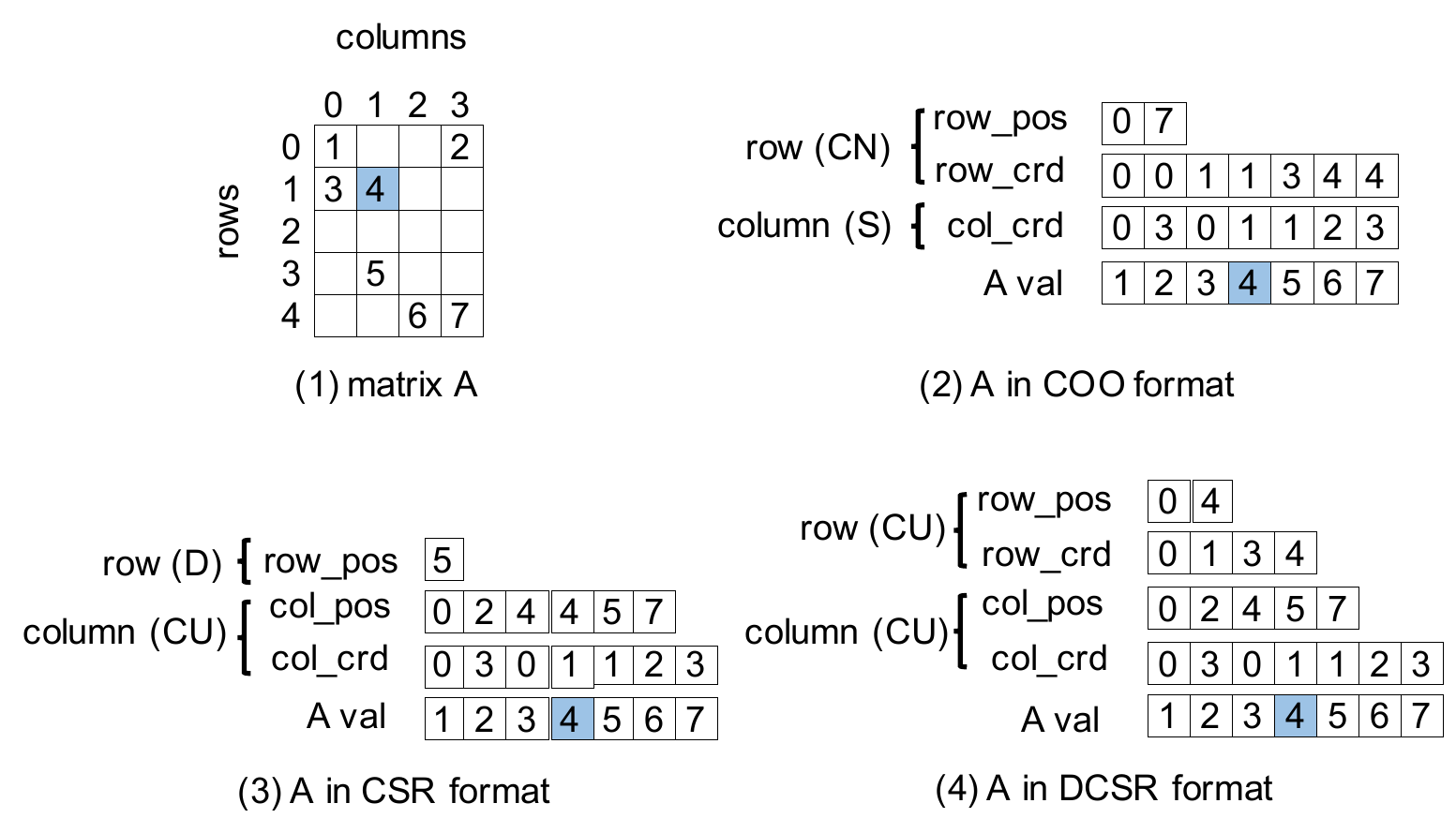}
        \label{fig:2dmatrix}
    }
    \hfill
    \subfloat[Sparse Tensor]{
        \includegraphics[width=0.48\textwidth]{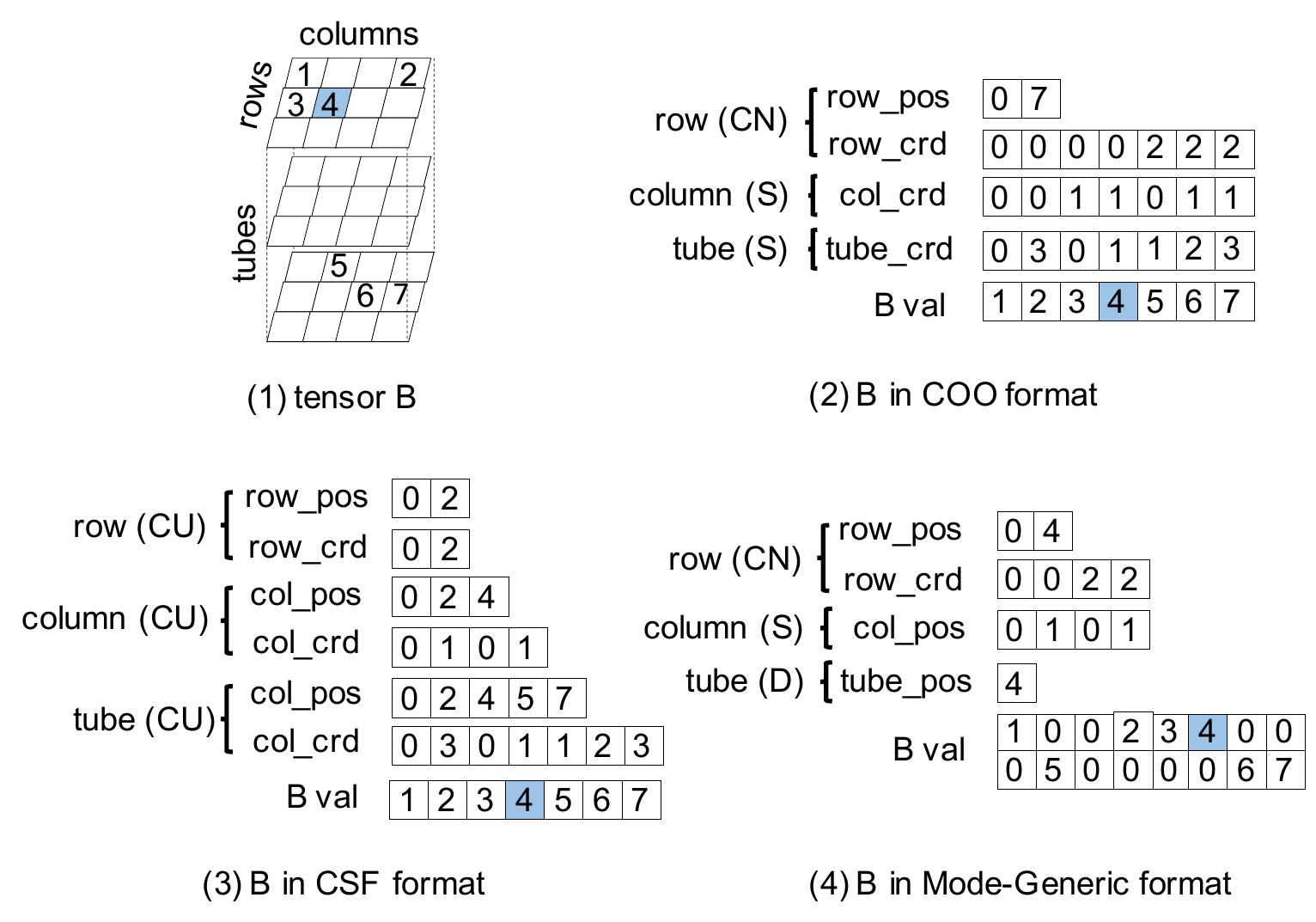}
        \label{fig:3dmatrix}
    }
    \caption{Example matrix and tensor represented in different formats. Each format is a combination of the storage format attributes.}
    \label{fig:attr_matrix}
\end{figure*}

\section{Tensor Storage Format}
\label{sec:format}

As reported in Section~\ref{sec:background}, to support multiple sparse storage formats a compiler needs a uniform way to represent each tensor in memory. This internal storage formats need to preserve the characteristics of the original format, e.g., data compression or performance for specific sparse patterns, while allowing a unified algorithm to generate efficient code for each computational expression.
\name{} defines a set of storage format attributes for each dimension to represent various sparse tensor formats.
Code generation is then based on each dimension's storage format attributes rather than the whole format, which greatly reduces the number of formats and combinations that a compiler needs to support.
Importantly, \name{} does not convert the original data layout into a different storage format. Instead, the storage format attributes are used to compose meta-data information that describes the original format, i.e., the data layout of the original format is preserved in memory and retains the original characteristics (compression, locality, etc.).

Representing every tensor dimension separately has been shown to be an effective way to \emph{generalize} tensor storage formats and support efficient code generation~\cite{kjolstad2017tensor}. Representing each dimension independently makes it easier to manage, adapt, and convert formats and to generate computational kernels uniformly.
\name{} defines the following four storage format attributes borrowed  from~\cite{smith2015splatt,kjolstad2017tensor,li2018hicoo}: 

\begin{itemize}
    \item[] \textbf{Dense (D).} This dimension is in the ``dense'' format, i.e., all  coordinates in this dimension will be accessed during the computations. For this format, we only use one scalar number stored in the \texttt{pos} array to represent the size of this dimension, such as the \texttt{row} dimension in Figure~\ref{fig:2dmatrix}(3). 
    \item[] \textbf{Compressed\_Unique (CU).} This dimension is in a ``compressed unique" format, i.e., the coordinates of nonzero elements in this dimension are compressed, and only the unique (no duplication) ones are stored in the array \texttt{crd}. It uses another array \texttt{pos} to store the start position of each unique coordinate, such as the \texttt{row}  dimension Figure~\ref{fig:2dmatrix}(4), where the elements 1 and 2 are in the same row, but only one row coordinate is stored in \texttt{row\_crd} array.
    \item[] 
    \textbf{Compressed\_Nonunique (CN).} This dimension is in a ``compressed non-unique" format, i.e., all the coordinates of nonzero elements will be recorded in \texttt{crd} array, and every coordinate in the \texttt{crd} array will be accessed one by one. CN then stores the start and the length of the \texttt{crd} array to the \texttt{pos} array, such as the \texttt{row} dimension Figure~\ref{fig:2dmatrix}(2), where all the row coordinates of the nonzeros are stored in \texttt{row\_crd} array, \texttt{row\_pos} only stores the start and the length of the \texttt{row\_crd} array. 
    \item[] \textbf{Singleton (S)}. The dimension is in a ``singleton" format, i.e., all the nonzero coordinates are recorded to the array \texttt{crd} without any other information, such as the \texttt{column} dimension Figure~\ref{fig:2dmatrix}(2), only the column coordinates of the nonzeros are stored in \texttt{row\_crd} array.
\end{itemize}



Internally, each tensor dimension is described by two arrays, a position (\texttt{pos}) and a coordinate (\texttt{crd}) array. \textbf{D} only uses the \texttt{pos} array to store the size of the dimension; the compressed storage format attributes \textbf{CU} and \textbf{CN} use both \texttt{pos} and \texttt{crd} arrays to store the nonzero coordinates and their positions;
\textbf{S} only uses the \texttt{crd} array to store the nonzero coordinates in the dimension.

Furthermore, Figure~\ref{fig:attr_matrix} shows two examples that store a sparse matrix and a sparse tensor, respectively, in three formats (COO, CSR, and DCSR) with the representation of varied storage format attributes combinations.
By properly combining the tensor storage format attributes, \name{} can represent the important sparse storage formats, including COO, CSR, DCSR, BCSR, CSB, ELLPACK, CSF and Mode-generic, in a uniform way, while retaining each format's characteristics.

\input{figure-formats-attr}

%% file: language.tex
\section{\name{} Language Definition}
\label{sec:language}

\name{} provides a high-level Tensor Algebra \gls{DSL} that increases portability and productivity by allowing scientists to reason about their algorithms implementation in their familiar notation and syntax.
Specifically, \name{} \gls{DSL} allows scientists 1) to express concepts and operations in a form that closely resembles their familiar notations and 2) to convey domain-specific information to the compiler for better program optimization. 
For example,
our language represents Einstein mathematical notation and provides users with an interface to express tensor algebra semantics.
The same \name{} program can be lowered to different architectures, and the lowering steps can follow different optimizations and lowering algorithms, allowing \name{} to produce high-quality code for target architectures without excessive burden on the programmer (see Section~\ref{sec:codegen}).
This work extends the \name{} tensor algebra language to support sparse tensor algebra operations and syntax, the storage formats described in the previous sections.



Furthermore, we extend \name to support dynamic data types.
As discussed above, Figure~\ref{fig:example_program} shows an example of a \name{} program. In the \name{} language, a tensor object refers to a multi-dimensional array of arithmetic values that can be accessed by indices. 
Range-based index label constructs (\texttt{IndexLabel}) represent the range of indices expressed through a scalar, a range, or a range with increment. Index labels can be used both for constructing a tensor or for representing a tensor operation. 
Different from the original \name{} compiler~\cite{blind}, \texttt{IndexLabel}s can now be defined as \emph{static} or \emph{dynamic}. Static \texttt{IndexLabel}s explicitly state the size of the dimension (Line 5) while dynamic \texttt{IndexLabel}s (Lines 3 and 4) only indicate that there exists a dimension, but the size will be determined later on during the execution of the program. Dynamic and static index labels differ in that dynamic index labels indicate an unknown size through a question mark (\texttt{?}) operator while static index labels explicitly state the size of the dimension through a scalar value.

A tensor is constructed by defined static or dynamic index labels and by declaring the sparsity of each dimension, according to the internal storage format described in the previous section. In Figure~\ref{fig:example_program} tensor \texttt{A} is stored in CSR format, while tensors \texttt{B} and \texttt{C} are stored in dense format. Note that \name{} provides convenient notation to represent the most common tensor storage format, avoiding the need to specify the storage format for each dimension, as described in the comments at Lines 8-10. Internally, however, \name{} reasons in terms of sparsity on each dimension when generating code.

In the example \name program in Figure~\ref{fig:example_program}, the tensor \texttt{A}, \texttt{B}, and \texttt{C} are initialized with a tensor file by \texttt{space\_read()}, the constant value \texttt{1.0}, and the constant value \texttt{0.0}, respectively. 
The function \texttt{space\_read()} first reads a tensor from the file in \texttt{COO} format and then converts it to our internal storage format (see Section~\ref{sec:format}) to represent \texttt{CSR}.
We implement \texttt{space\_read()} as a \emph{runtime function}, and it can be called in the \name program directly.

The last line in the program performs the SpMM operation. However, users need not explicitly state that the operation is an SpMM but can simply use the common tensor contraction \texttt{*} operator. \name{} will infer that the operator refers to an SpMM operation from the storage format of the input tensors, in this case, a sparse matrix and a dense matrix, and will generate the proper code to iterate over the specific storage format through rules generated from the definition of storage format attributes.
Also, note that \name{} employs index labels to determine the type of operation to perform. For example, the \texttt{*} operator refers to a tensor contraction if the contraction indices are adjacent or to element-wise operation otherwise.
In Figure~\ref{fig:example_program}, the index label \texttt{b} is used as contraction indices between \texttt{A} and \texttt{B} (adjacent or internal indices), thus the operator \texttt{*} refers to a tensor contraction.
Therefore, \emph{\name can not only support tensor contraction but are generally applicable to many other operations as well.}
Conclusively, the \name{} TA language simplifies writing tensor algebra program by supporting common programming paradigms and enables users to express high-level concepts in their familiar notations.

%% file: codegen.tex
\section{Compilation Pipeline}
\label{sec:codegen}




We introduce sparse tensor algebra dialect in \gls{MLIR} to support mix dense/sparse tensor algebra computation with a wide range of storage formats.
We use format attributes to represent each dimension sparsity format in a uniform way in the proposed TA \gls{IR}. \name compiler generates efficient code based on the represented format attribute per dimension. This section describes the compiler framework, which consists of two main parts: 1) a sparse \gls{MLIR} TA dialect to represent tensor storage formats and operations, and 2) code generation algorithms to generate efficient serial and parallel code starting from the proposed \gls{TA} DSL.  


\begin{figure}[t]
  \begin{center}
    \includegraphics[width=0.5\textwidth]{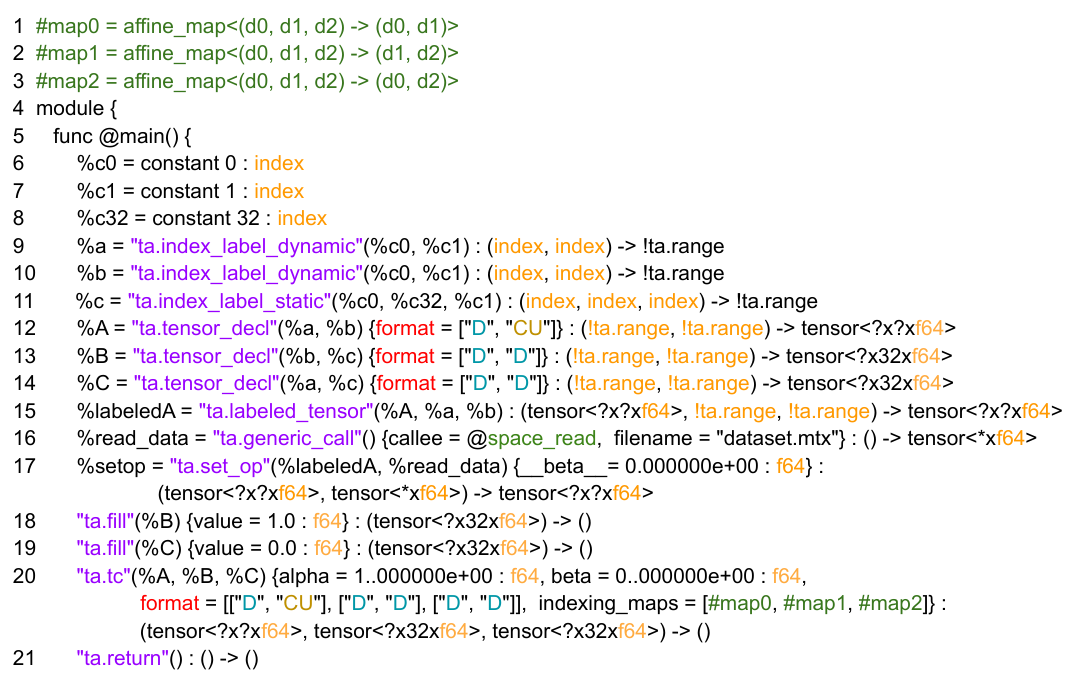}
  \end{center}
  \caption{Generated sparse tensor algebra dialect for SpMM operation}
  \label{fig:spdialect}
\end{figure}

\subsection{Sparse Tensor Algebra Dialect}

\name{ }supports a uniform tensor storage format based on the attributes described in Section~\ref{sec:format} and the tensor algebra operations supported in our \gls{DSL}. 
Figure~\ref{fig:spdialect} shows the generated tensor algebra \gls{IR} for the SpMM program in Listing~\ref{fig:example_program}.
The rest of this section details the various operation in the sparse TA dialect.


\textbf{Static/Dynamic Index Labels.}
The sparse tensor algebra dialect supports two types of index label, static and dynamic. If the dimension size of the index is known in compile-time, \name{} uses {\tt ta.$index\_label\_static$} to represent the index label. It has three operands, which represent the start, end, and step value on this index.  {\tt ta.$index\_label\_dynamic$} is used to represent the index label when the dimension size is unknown in compile time. {\tt ta.$index\_label\_dynamic$} has two operands, the start, and step value on this index. The end value on this index will be known in runtime. 

\textbf{Sparse Tensor Declaration.}
In sparse tensor algebra dialect, the tensor is declared with {\tt ta.$tensor\_decl$} operation. The operands of {\tt ta.$tensor\_decl$} are the index labels of the tensor. It can contain an arbitrary number of operands, which means it can declare arbitrary dimensional tensor. {\tt ta.$tensor\_decl$} operation also contains storage format attributes of the tensor in each dimension for sparse tensors.

\textbf{Sparse Tensor Operations.} The sparse TA dialect also defines the tensor algebra operations supported by \name{}. For example, the tensor contraction \texttt{ta.tc} operation for an SpMM computation (shown in line 20 of Figure ~\ref{fig:spdialect}) takes two input tensors and computes the result of the contraction. The first and second operands (\%A and \%B) are input tensors, and the third operand (\%C) is the output tensor. ``ta.sptensor$<$tensor$<$?$\times$i32$>$, tensor$<$?$\times$i32$>$, tensor$<$?$\times$i32$>$, tensor$<$?$\times$i32$>$, tensor$<$?$\times$f64$>>$'' 
is the data type for \%A, while 
``tensor<?$\times$32$\times$f64>'' is the data type for \%B, and ``tensor$<$?$\times$32$\times$f64$>$'' is the data type for \%C. 
``-> ()'' represents the return type which is \texttt{void}.

We introduce \texttt{formats} attribute to extend the original \texttt{ta.tc} to provide the storage format information of each input tensor.
In line 20 of Figure~\ref{fig:spdialect}, the first tensor is in CSR format, while the second and third are all Dense tensors. 
The code in the figure shows that each input tensors is associated with its storage format information. 
We also introduce \texttt{indexing\_maps} to
\texttt{ta.tc} to represent the indices of each tensor. 
The \texttt{indexing\_maps} helps propagate indices information along with the lowering stack.
The tensor expression and the storage format information will be further propagated down to the lower level of the \gls{IR} to provide the format attribute in each dimension when generating the computational code. 


\begin{figure}[t]
  \begin{center}
    \includegraphics[width=0.48\textwidth]{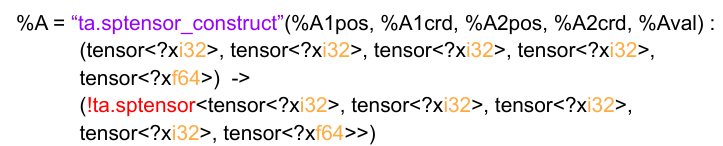}
  \end{center}
  \caption{Sparse tensor data structure construction operation}
  \label{fig:sptensor}
\end{figure}

\textbf{Sparse Tensor Data Type.} As described in Section ~\ref{sec:format}, a tensor $T$ consists of $k$ dimensions $d_i$ for $0 \leq i\leq k-1$, where every dimension $d_i$ is associated with a uniform storage attribute $a_i \in \{\textbf{D}, \textbf{CU}, \textbf{CN}, \textbf{S}\}$. \name{} associates two arrays \texttt{crd} and \texttt{pos} to each dimension to describe the storage format (meta-data). 
In the TA dialect, we define a sparse tensor as a \texttt{struct} 
data structure, which contains the nonzero indices in each dimension and their values. 

Figure ~\ref{fig:sptensor} shows how a 2D sparse matrix is represented in our TA dialect. In Figure~\ref{fig:sptensor},  \texttt{ta.sptensor\_construct} is the function to construct the sparse tensor struct, which is implemented as an operation in the TA dialect.
The \texttt{sptensor\_construct} operation takes the \texttt{pos} and \texttt{crd} arrays in each dimension (\%A1pos, \%A1crd, \%A2pos, \%A2crd) and the nonzero values (\%AVal) as input, and returns a \texttt{ta.sptensor} type 
data structure that represents a sparse tensor in the TA dialect. 
The tensor types within \texttt{ta.sptensor} represent the 
\texttt{pos} and \texttt{crd} arrays corresponding to each dimension of the tensor itself (see Section~\ref{sec:format}).
In the \texttt{ta.sptensor} structure, the type of \%A1pos, \%A1crd, \%A2pos, \%A2crd are tensor$<$?$\times$i32$>$, the type of \%Aval is tensor$<$?$\times$f64$>$.

\subsection{Sparse Code Generation Algorithm}
\label{sec:codegen_alg}


\name{} lowers the code from high-level \name{} \gls{DSL} language to low-level machine code in multiple lowering steps. 


\textbf{DSL Lowering.} The first step in our compilation pipeline consists of lowering the high-level \name{} \gls{DSL} into the sparse TA dialect. 
Figure~\ref{fig:spdialect} shows the TA dialect corresponding to the \name{} code presented in Figure ~\ref{fig:example_program}. 
In Figure~\ref{fig:spdialect}, "\texttt{ta.}" represents the tensor algebra dialect. 
The \texttt{indexLabel} operation in \name{} DSL will be lowered either into a \texttt{ta.index\_label\_static} operation or a \texttt{ta.index\_label\_dynamic} operation (e.g., Lines 9-11 in Figure~\ref{fig:spdialect}) based on whether the size of the dimension represented by the index label is known or unknown at compile time. 
The \texttt{ta.index\_Label} operation has three parameters (\%A, \%a, and \%b), which are the start, the end, and the iteration step values in the dimension represented by the index label.
The \texttt{IndexLabel} at Lines 3-4 of Figure ~\ref{fig:example_program} has an unknown size, so it will be lowered into the \texttt{ta.index\_label\_dynamic} operation, which only contains the start value of the dimension. The dimension size 
will be inferred during the runtime. 

\textbf{Progressive Lowering.} Next, the sparse TA dialect is further translated to lower \gls{MLIR} dialects. 
We describe this lowering process in two parts, early lowering and late lowering . 


First, in the early lowering step, \name{} lowers 
all the operations in the sparse TA dialect, except the \texttt{ta.tc} operation.
In particular, the \texttt{ta.tensor\_decl} operation, which declares a tensor, is lowered into \texttt{alloc} and \texttt{tensor\_load} operations, which are standard dialect operations in \texttt{std} dialect for dense. 
For sparse, \texttt{ta.tensor\_decl} operations are lowered into more, a composition of \texttt{alloc} and \texttt{tensor\_load} operations for \texttt{pos} and \texttt{crd} arrays to store the coordinates of nonzeros in each dimension, and \texttt{val} array to store nonzero values. 
These coordinates of nonzeros are later used by \texttt{ta.sptensor\_construct} operation (Figure~\ref{fig:sptensor}) to construct a sparse tensor. 
To fill the \texttt{pos}, \texttt{crd}, and \texttt{val} arrays, the \texttt{ta.generic\_call} operation is invocated to to call the \texttt{space\_read()} function.
The \texttt{ta.generic\_call} operation is then lowered to the \texttt{call} operations in the MLIR \texttt{std} dialect. 
The \texttt{ta.fill} operation initializes dense tensors with identical values. 
The \texttt{ta.fill} operation will be lowered into the \texttt{fill} operation in the MLIR $linalg$ dialect.
The \texttt{ta.return} operation returns the function, and is lowered into \texttt{return} operation in the MLIR $std$ dialect. 
The \texttt{ta.index$\_$label$\_$dynamic} operations is lowered into the \texttt{ta.index$\_$label$\_$static} operation when the index label is identified from the input file. 

\begin{figure}[t]
  \begin{center}
    \includegraphics[width=0.49\textwidth]{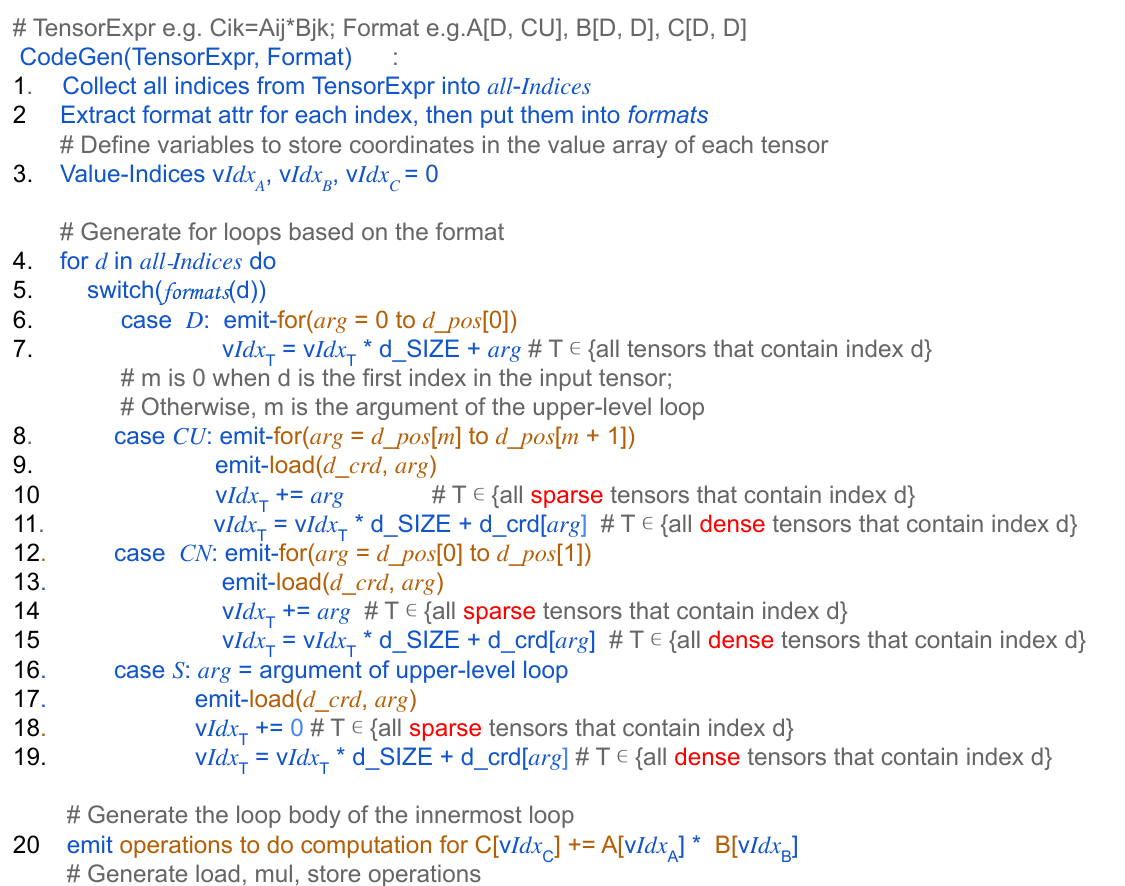}
  \end{center}
  \caption{Sparse code generation algorithm}
  \label{fig:codegen}
\end{figure}

Second, in the late lowering step, \texttt{ta.tc} operations are lowered into the MLIR \texttt{scf} (structure control flow) dialect operations. 
Figure~\ref{fig:codegen} describes the lowering algorithm to \texttt{ta.tc} with an example mix sparse dense tensor contraction operation, 
where a sparse tensor $A$ times a dense tensor $B$, and the output can be either sparse or dense.
The algorithm takes 
\texttt{ta.tc} as input, 
and automatically generates the computational kernel code of a combination of \texttt{scf} and \texttt{std} dialects. 
\texttt{ta.tc} is the sparse tensor algebra dialect of the tensor contraction operation presented at Line 20 in Figure~\ref{fig:example_program}.
As shown at Line 20 in Figure~\ref{fig:spdialect} 
\texttt{ta.tc} operation is lowered based on the code generation algorithm in Figure~\ref{fig:codegen}.


Figure~\ref{fig:codegen} shows \name{}'s code generation algorithm that consists of three key steps. 
This algorithm is general, applicable to varied tensor algebra operations, and can generate arbitrary index permutations. Moreover, in contrast to TACO, \name{} can generate sparse output.
Take tensor expression $C_{ik} = A_{ij} * B_{jk}$ as an example, and assume the format of $A$ is [\textbf{D}, \textbf{CU}], $B$ is [\textbf{D}, \textbf{D}] and $C$ is [\textbf{D}, \textbf{D}], respectively. The basic idea of this code generation is as follows:


{\bf Step-I} (Line 1 to Line 3) collects both index information as well as the format attribute of each index. The above sample tensor expression has three indices ($all$-$Indices = \{i, j, k\}$). The order of these indices matters, and is decided by tensor access orders. The format attribute of each index is decided by the usage of this index. If this index appears in dense input tensors only, its format attribute is \textbf{D}; otherwise, the format attribute is decided by the corresponding dimension of the sparse tensor. For the above sample tensor expression, the format attribute of index $i$ is \textbf{D} and $j$ is \textbf{CU} (both decided by sparse input tensor $A$), and $k$ is \textbf{D} (decided by dense input tensor $B$), respectively. After collecting this information, this algorithm defines three index variables ($vIdx_A$, $vIdx_B$ and $vIdx_C$) to access the value array of tensor $A$, $B$ and $C$, respectively (Line 3).  

{\bf Step-II} (Line 4 to Line 19) iterates each index to generate loop structure code (as the algorithm line starting with "emit" shows). It leverages the aforementioned definition of each storage format attribute to find nonzero coordinates in each dimension via \texttt{pos} and \texttt{crd} arrays (e.g., $d\_pos$ and $d\_crd$ in the algorithm). 
Table~\ref{table:format-code} shows the sample loop code in C language for each format attribute. 
Besides generating loop structure code for each index, this step also updates three index variables ($vIdx_T$, $T \in \{A, B, C\}$) that will be used for inner-most computation. If the format attribute of an index (e.g., $d$) is $D$, i.e., $d$ only appears in dense tensors, then $vIdx_T = vIdx_T \times d\_SIZE + arg$, where $T$ denotes all dense tensors that contain index $d$, $arg$ is the coordinate on index $d$ (i.e., the argument of the generated loop for index $d$), and $d\_SIZE$ is 
index $d$'s dimension size. If the format attribute of index $d$ is sparse (e.g., $CU$), this step handles sparse tensors and dense tensors separately. For sparse tensors $T$ that contain index $d$, $vIdx_T = vIdx_T + arg$, where $arg$ is still the argument of the generated loop for index $d$. 
For dense tensors $T$ that contain index $d$, $vIdx_T = vIdx_T + d\_crd[arg]$, where $d\_crd$ is the \texttt{crd} array of index $d$, and $d\_crd[arg]$ is the coordinate.



{\bf Step-III} (Line 20) generates inner-most computation code to load values from $A[vIdx_A]$ and $B[vIdx_B]$, compute their product, and update $C[vIdx_C]$, after step-II generates $vIdx_T$ for tensor $T$ ($T \in \{A, B, C\}$). 

\begin{table}
\caption{\label{table:format-code}Generated code to access nonzeros coordinates}
\scriptsize 
\begin{tabular}{p{0.5cm}|p{6.5cm}}
 \hline
 Attr & Corresponding code \\ [0.5ex] 
 \hline\hline
 \textbf{D} & \texttt{for i from 0 to pos[0] \{ ... \} } \\ 
 \hline
 \textbf{CU} & 
 \texttt{for i from pos[m] to pos[m+1]\{ idx = crd[i];\} 
 }\newline (m: The argument of the upper level loop. m is 0 when the dimension is the first dimension of the tensor) 
 \\
 \hline
 \textbf{CN} & \texttt{for i from pos[0] to pos[1]\{idx = crd[i];\} } \\
 \hline
 \textbf{S} & \texttt{idx = crd[m]; } \\
 \hline
\end{tabular}
\end{table}

\begin{figure}[t]
    \includegraphics[width=0.42\textwidth]{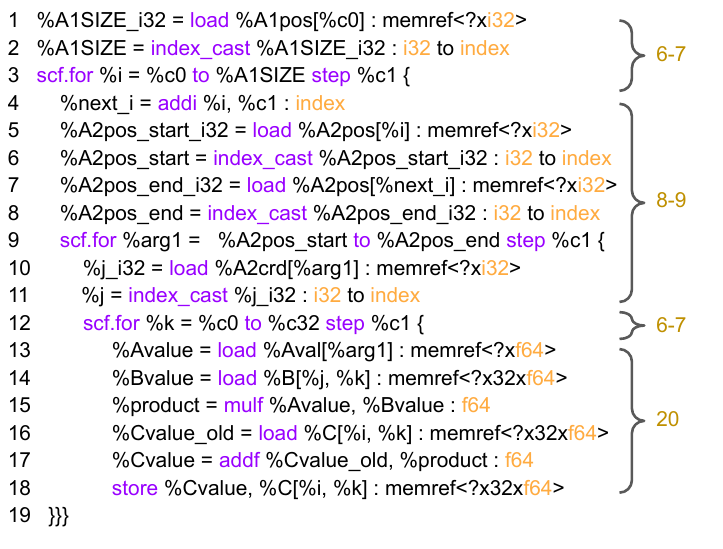}
  \caption{Lowered \texttt{scf} dialect code example for SpMM in the CSR format. The right side numbers represent line numbers in Algorithm~\ref{fig:codegen}} 
  \label{fig:spmv-csr_new}
\end{figure}

\subsection{Parallel Code Generation}
For sequential execution \name{} lowers the \texttt{scf} dialect to the \texttt{llvm IR} dialect and then to proper LLVM \gls{IR} for assembly and linking. For parallel execution, instead, the \texttt{scf} dialect is lowered to the \texttt{async} dialect (See Figure~\ref{fig:overview}). In details, we developed a pass to lower \texttt{scf.for} loops to \texttt{scf.parallel} loops and the latter to the \texttt{async} dialect. 
The \texttt{async} dialect encapsulates the semantics of an asynchronous task-based parallel runtime in which computational tasks are spawn and asynchronously executed by parallel worker threads. Currently, \gls{MLIR} supports a task continuation stealing approach (like Cilk~\cite{blumofe1995cilk}) in which the control is returned to the parent task after spawning. The dialect provides semantics primitives to synchronize the execution of tasks. \name{} lowers those asynchronous tasks execution primitives to LLVM co-routines in LLVM IR, which is then passed to the assembler and linker to create a binary.
As Figure~\ref{fig:spmm_multithreaded} shows, the \gls{MLIR} asynchronous runtime introduces relatively low overhead during execution, which improves performance, especially for small computations.

%% file: opt.tex
\section{Data Reordering}
\label{sec:opt}



The distribution of the nonzero entries in sparse matrices/tensors can significantly affect the performance of sparse matrix/tensor algebra computations. 
Reordering~\cite{jiang2020novel, li2019efficient} is the de facto technique to optimize the memory access pattern caused by uneven data distribution. 
Different from existing compiler frameworks~\cite{kjolstad:2017:tacotool,kiriansky2016optimizing} which apply reordering to iterations, we apply reordering to matrices and tensors to optimize their memory access patterns.

We borrow from the reordering algorithm presented in~\cite{li2019efficient} (\texttt{LexiOrder}),  extended it to support sparse matrices, and implemented it in the \name{} runtime (\texttt{tensor\_reorder()}). 
The \texttt{LexiOrder} algorithm is built on top of the doubly lexical ordering algorithm~\cite{lubiw1987doubly,paige1987three} with some optimization techniques to advance its overall efficiency and availability on some concern cases. 
The basic idea of the \texttt{LexiOrder} algorithm is to sort a specific dimension (either rows or columns for matrices) in an iteration using the doubly lexical ordering algorithm and sort all dimensions in turn across iterations. 
The algorithm's objective is to cluster 
all nonzero entries 
around the diagonal to increase spatial and temporal locality. 

%% file: evaluation.tex
\label{sec:evaluation}
\input{ops-formats}

\section{Evaluation}
In this section we evaluate \name{} against state-of-the-art high-level compiler frameworks and \gls{DSL} for dense and sparse tensor algebra. Specifically, we compare our results against TACO~\cite{kjolstad2017tensor}, a tensor algebra compiler that performs automatic source-to-source transformation from TACO \gls{DSL} to sequential C++, Parallel OpenMP, and data-parallel CUDA. 
For brevity, we evaluated the performance of selected benchmarks with a single storage format -- matrices (CSR) and tensors (CSF), though our compiler can operate on other formats as well.
All results reported are the average of 25 runs.

\subsection{Experimentation Setup}

We performed our experiments on a compute node equipped with two Intel Xeon Gold 6126 sockets running at 2.60GHz. Each CPU socket consists of 12 processing core (for a total of 24 cores). The system features 192 GB of DRAM memory.
We compiled \name{}, TACO, and all the benchmarks with $-$\texttt{O3} and \texttt{clang 12.0} and use the most recent MLIR version at the time of writing this manuscript. 

We use as input datasets $2833$ matrices and six tensors of different sizes and shapes chosen from 
the SuiteSparse Matrix Collection~\cite{davis2011university}, the FROSTT Tensor Collection~\cite{frosttdataset}, and BIGtensor~\cite{haten2_ICDE2015}. 
The SuiteSparse Matrix Collection 
is a growing dataset of sparse matrices in real-world applications. 
The dataset is widely used in the numerical linear algebra community for performance evaluation.
The FROSTT Tensor Collection is a composition of open-source sparse tensor datasets from various data
sources that are difficult to collect. 
The BIGtensor dataset is a tensor database that contains large-scale tensors for large-scale tensor analysis. 
Our input datasets represent the most important HPC domains in scientific computing, including chemistry, structural engineering, various linear solvers, computer graphics and vision, and molecular dynamics.
We provide the description of the six tensors in Table~\ref{tab:matrix_tensor}.


\begin{table}
\footnotesize
\begin{tabular}{|p{1.2cm}|p{1.5cm}|p{1.2cm}|p{3cm}|}
\hline
Name & Size & Nonzeros & Domain 
\\ \hline
\hline 
NELL-1 & 2,902,330 x 2,143,368 x 25,495,389 & 143599552 & Natural Language Processing \\ \hline
NELL-2 & 12,092 x 9184 x 28,818 & 76879419 & Natural Language Processing \\\hline
delicious-3d &	532,924 x 17,262,471 x 2,480,308 & 140,126,181 & Tags from Delicious website\\ \hline
flickr-3d &	319,686 x 28,153,045 x 1,607,191 & 112,890,310 & Tages from Flickr website\\ \hline
vast-2015-mc1-3d & 165,427 x 11,374 x 2 & 26,021,854 & Theme park attend event\\ \hline
Freebase-music\cite{haten2_ICDE2015} &	23,344,784 x 223,344,784 x 166 & 99,546,551 & Entries related with music in Freebase\\ \hline
\end{tabular}
\caption{Description of sparse tensors}
\label{tab:matrix_tensor}
\end{table}


\subsection{Sparse Tensor Operations}

We define the sparse tensor operations considered in \name below.

\textbf{SpMV.}
The Sparse Matrix-times-Vector (SpMV or SpMSpV), $\V{Y} = \M{X} \times \V{V}$, is the multiplication of a sparse matrix $\M{X} \in \mathbb{R}^{I_1 \times I_2}$ with a dense vector $\V{V} \in \mathbb{R}^{I_2}$. 
$y_{i_1} = \sum_{i_2=1}^{I_2} x_{i_1 i_2} v_{i_2}$.

\textbf{SpMM.}
The Sparse Matrix-times-Matrix (SpMM or SpGEMM), $\M{Y} = \M{X} \times \M{U}$, is the multiplication of a sparse matrix $\M{X} \in \mathbb{R}^{I_1 \times I_2}$ with a dense matrix $\M{U} \in \mathbb{R}^{I_2 \times R}$. 
$y_{i_1 r} = \sum_{i_2=1}^{I_2} x_{i_1 i_2} u_{i_2 r}$.

\textbf{SpTTV.}
The Sparse Tensor-Times-Vector (SpTTV)~\cite{TTB_Sparse} in mode $n$, $\T{Y} = \T{X} \times_n \V{V}$, is the multiplication of a sparse tensor $\T{X} \in \mathbb{R}^{I_1 \times I_2 \times I_3}$ with a dense vector $\V{V} \in \mathbb{R}^{I_n}$, along mode $n$. 
Given $n=1$,
$y_{i_2 i_3} = \sum_{i_1=1}^{I_1} x_{i_1 i_2 i_3} v_{i_n}$.
This results in a two-dimensional $I_2 \times I_3$ tensor which has one less dimension.

\textbf{SpTTM.}
The Sparse Tensor-Times-Matrix (SpTTM)~\cite{kolda2009tensor,TTB_Sparse} in mode $n$,  denoted by $\T{Y} = \T{X} \times_n \M{U}$, is the multiplication of a sparse tensor $\T{X} \in \mathbb{R}^{I_1 \times I_2 \times I_3}$ with a dense matrix $\M{U} \in \mathbb{R}^{I_n \times R}$, along mode $n$. 
Mode-$1$ TTM results in a $R \times I_2 \times I_3$ tensor, and its operation is defined as
$y_{r \cdots i_2 i_3} = \sum_{i_n=1}^{I_n} x_{i_1 i_2 i_3} u_{i_n r}$.
Also, note that $R$ is typically much smaller than $I_n$ in low-rank decompositions, typically $R < 100$.

SpMV and SpMM widely appear in applications from scientific computing, such as direct or iterative solvers~\cite{lee2004performance,williams2007optimization}, to data intensive domains~\cite{zhang2009automatic}, graph analytics~\cite{li2013smat}.
SpTTV and SpTTM are computational kernels of popular tensor decompositions, such as the Tucker decomposition
~\cite{kolda2009tensor,yokota2014multilinear,sidiropoulos2017tensor}
, tensor power method
~\cite{anandkumar2017analyzing,wang2015fast}, for a variety of applications, including (social network, electrical grid) data analytics, numerical simulation, machine learning.

\subsection{Performance Evaluation}

\begin{figure*}[htbp]
\begin{subfigure}[b]{0.247\textwidth}
    \includegraphics[width=\textwidth]{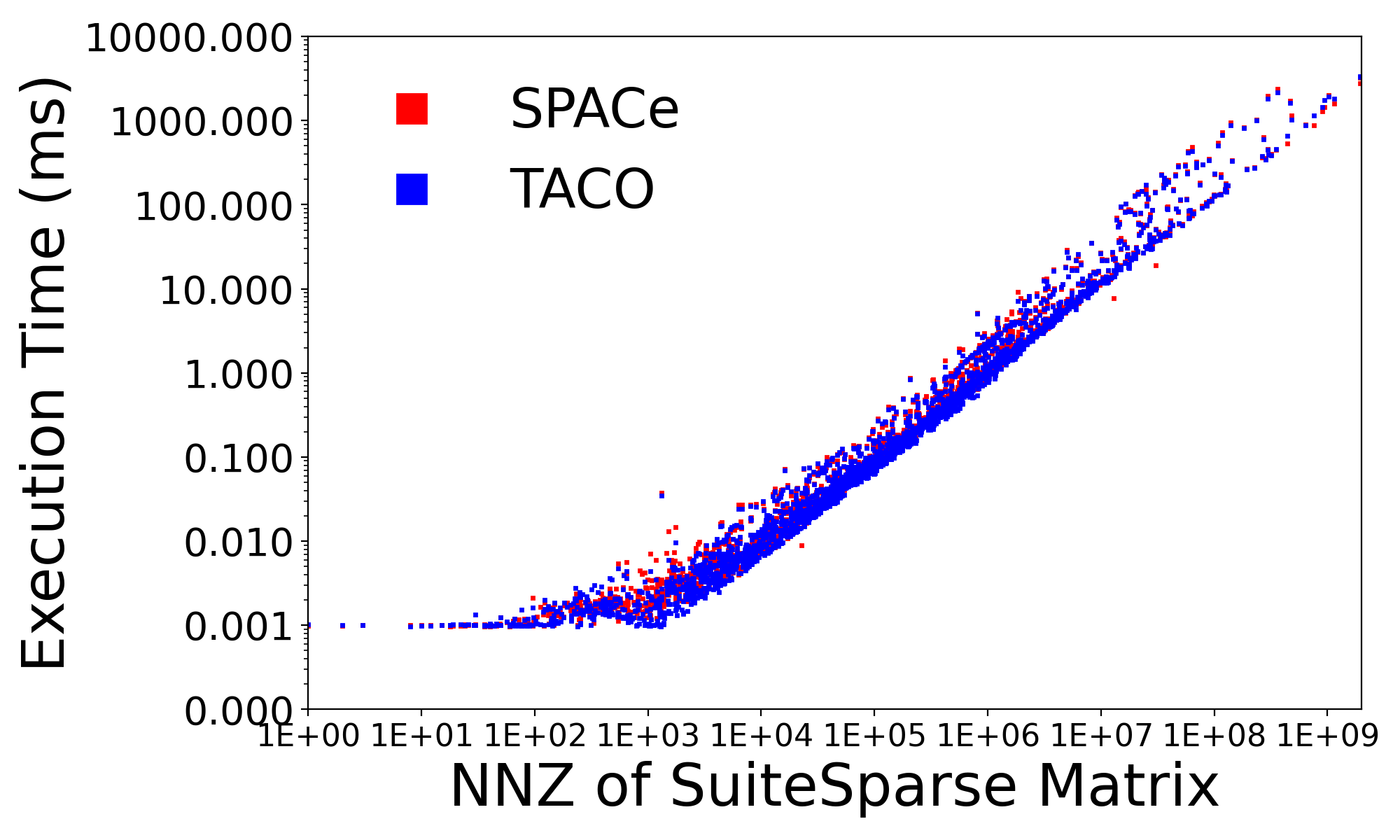}
    \caption{sequential SpMV}
    \label{fig:spmv_seq}
\end{subfigure}
\begin{subfigure}[b]{0.247\textwidth}
    \includegraphics[width=\textwidth]{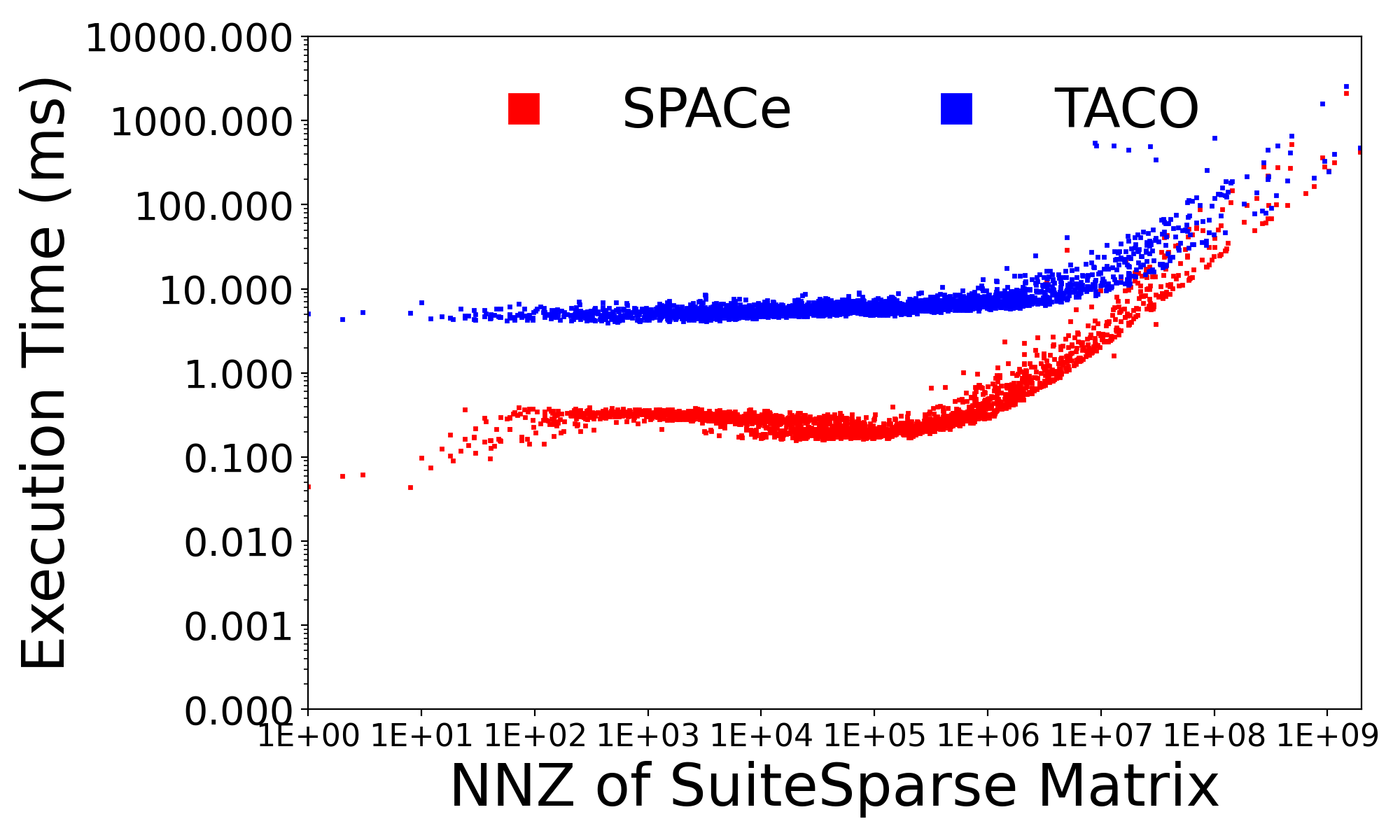}
    \caption{parallel SpMV}
    \label{fig:spmv_multithreaded}
\end{subfigure}
\begin{subfigure}[b]{0.247\textwidth}
    \includegraphics[width=\textwidth]{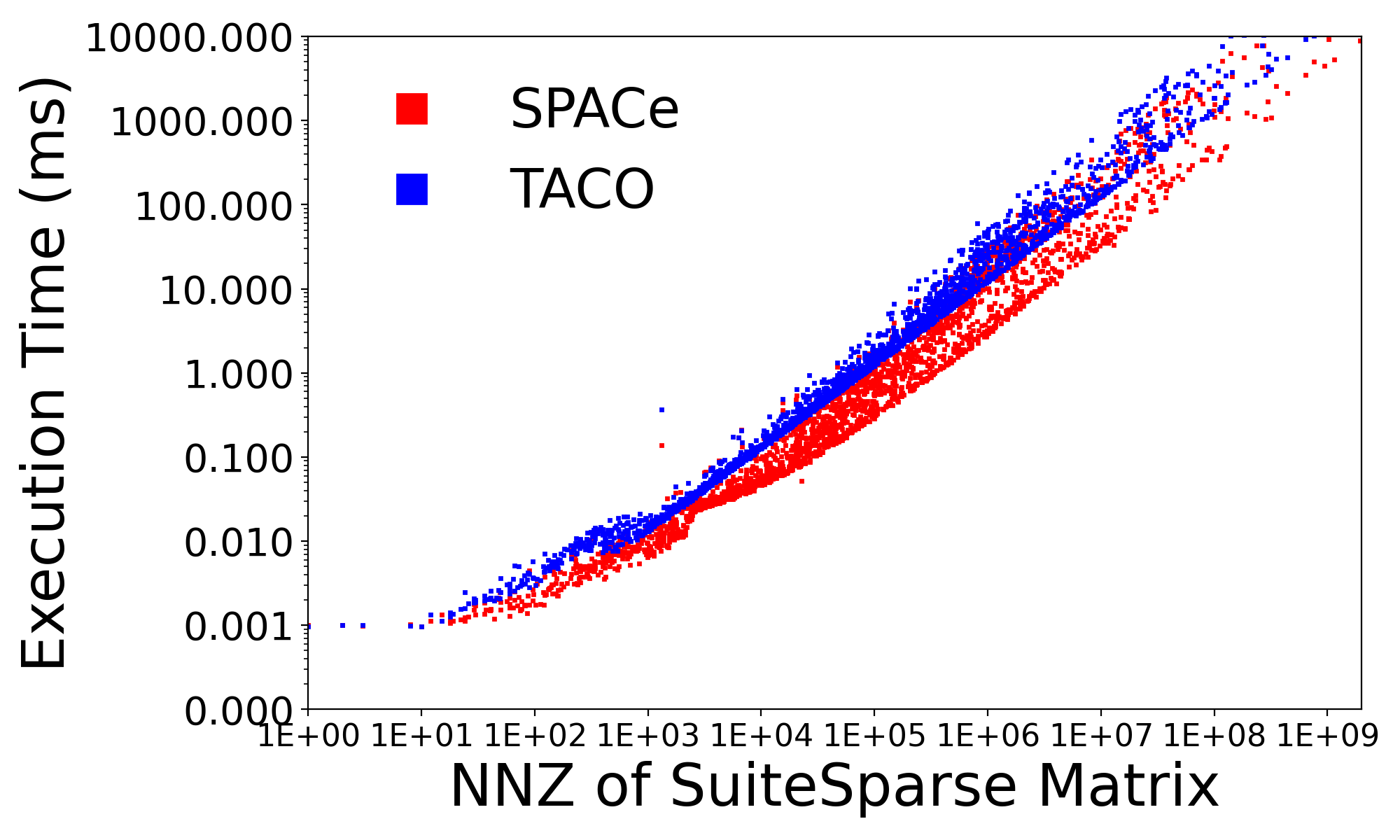}
    \caption{sequential SpMM} 
    \label{fig:spmm_seq}
\end{subfigure}
\begin{subfigure}[b]{0.247\textwidth}
    \includegraphics[width=\textwidth]{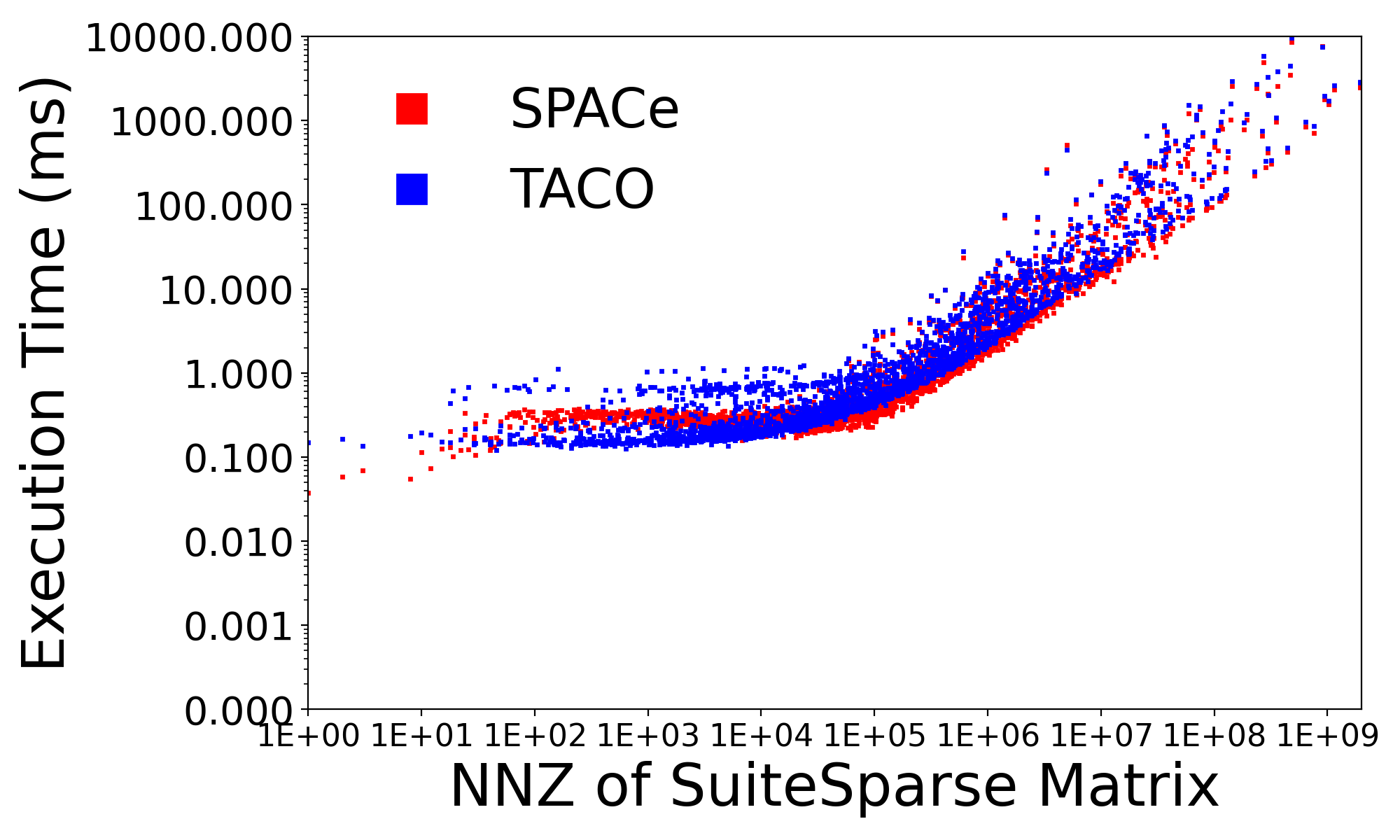}
    \caption{parallel SpMM}
    \label{fig:spmm_multithreaded}
\end{subfigure}
\caption{Performance comparison with TACO on CPU.}
  \label{fig:cpu_performance}
\end{figure*}

\begin{figure*}[htbp]
\begin{subfigure}[b]{0.247\textwidth}
    \includegraphics[width=\textwidth]{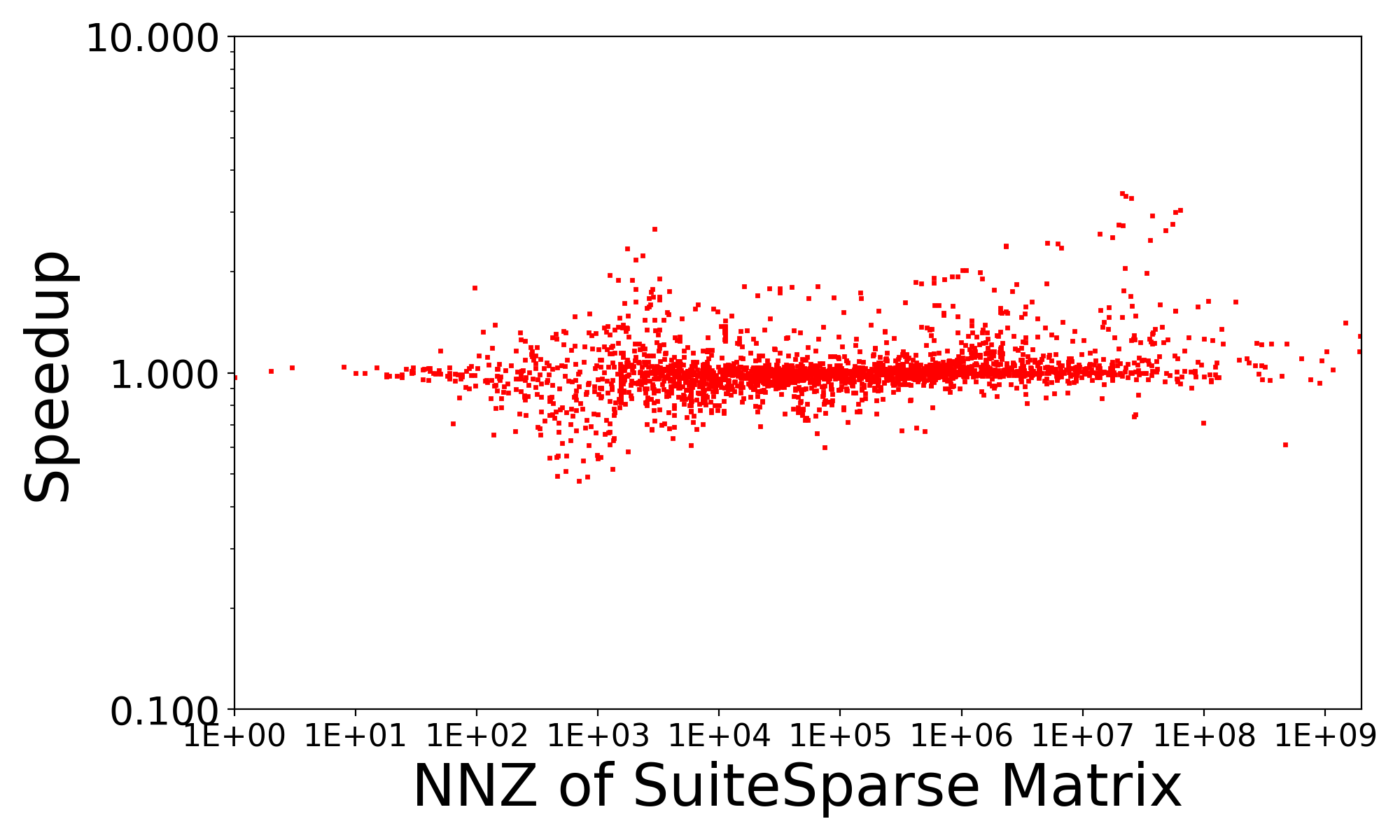}
    \caption{sequential SpMV-lexi}
    \label{fig:spmv_seq_ordering}
\end{subfigure}
\begin{subfigure}[b]{0.247\textwidth}
    \includegraphics[width=\textwidth]{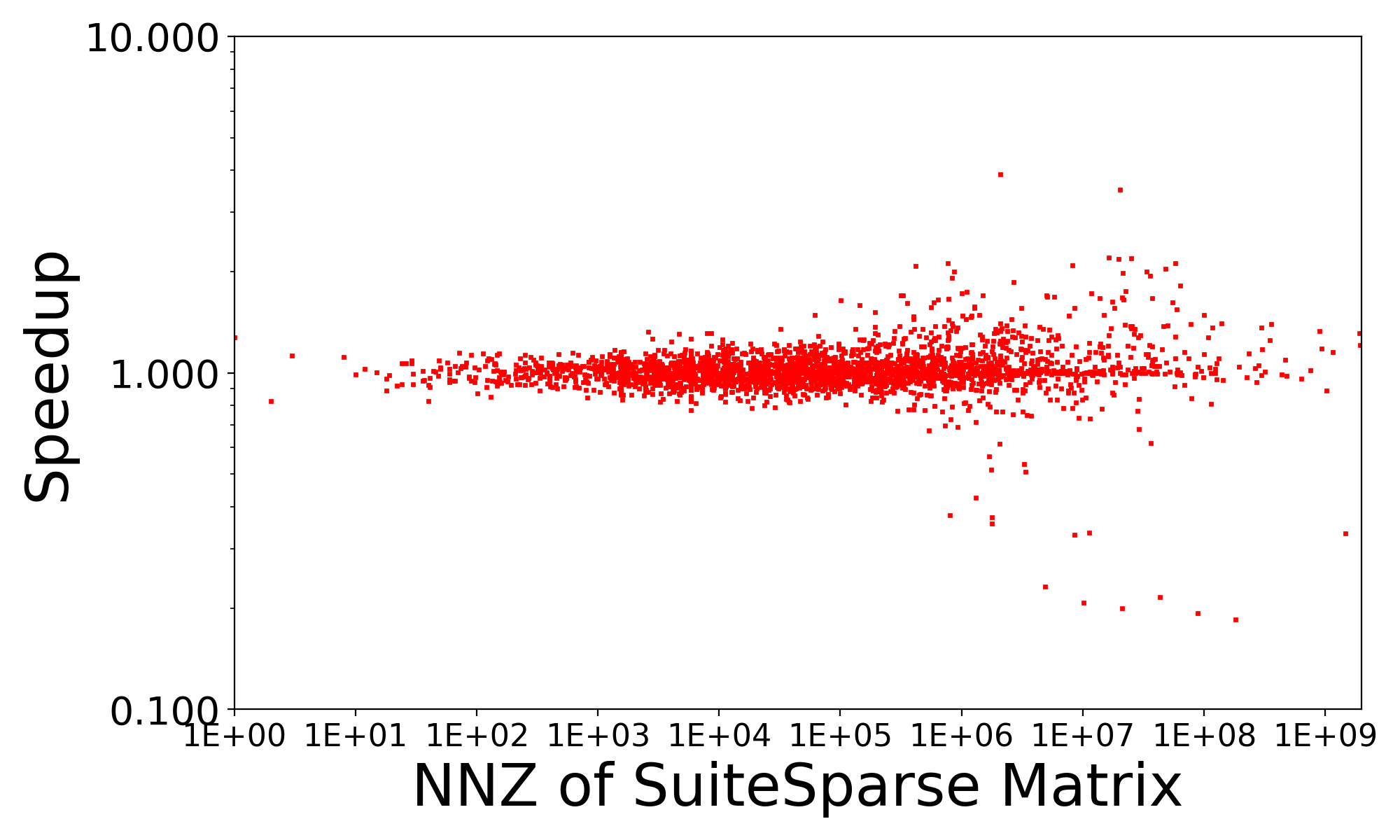}
    \caption{parallel SpMV-lexi}
    \label{fig:spmv_multithreaded_ordering}
\end{subfigure}
\begin{subfigure}[b]{0.247\textwidth}
    \includegraphics[width=\textwidth]{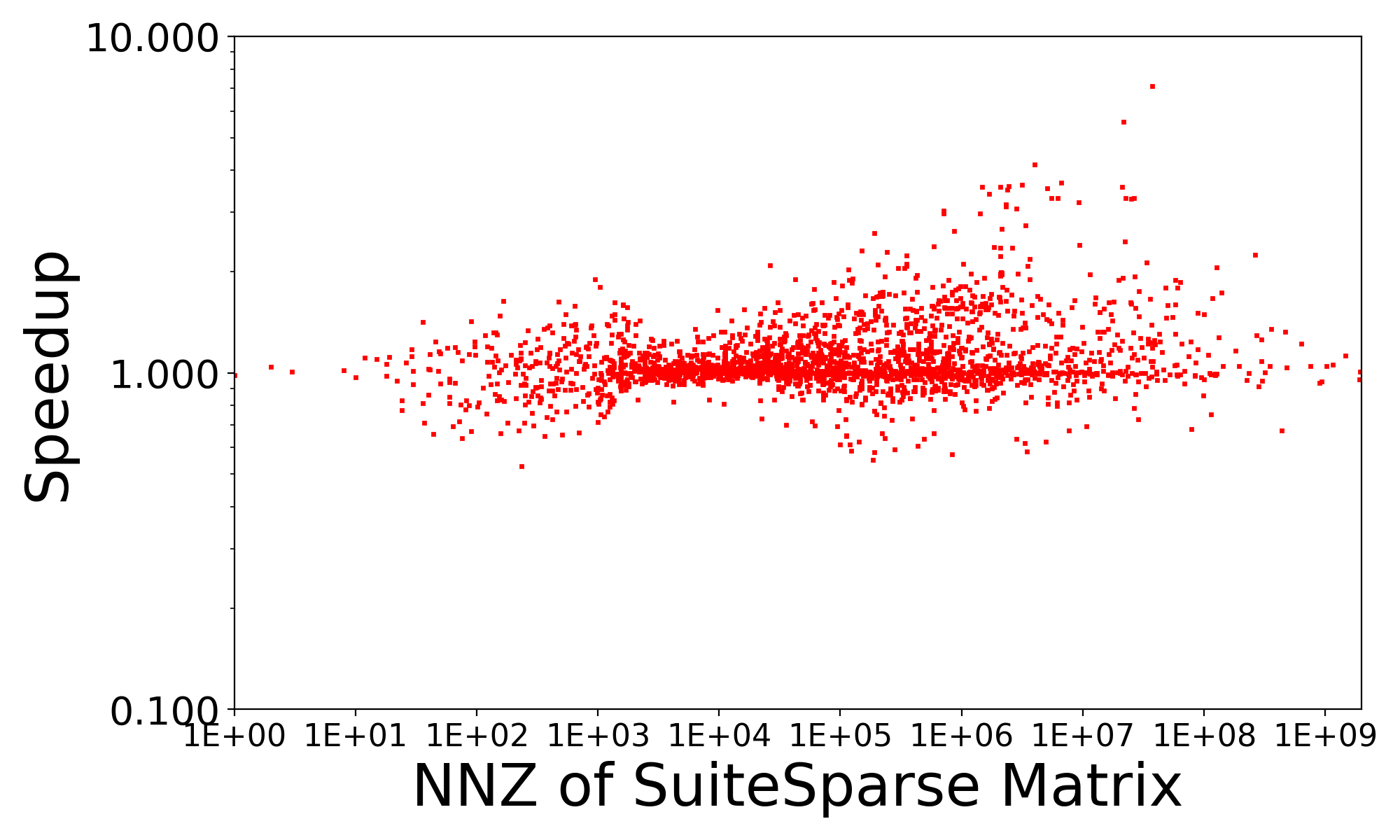}
    \caption{sequential SpMM-lexi}
    \label{fig:spmm_seq_ordering}
\end{subfigure}
\begin{subfigure}[b]{0.247\textwidth}
    \includegraphics[width=\textwidth]{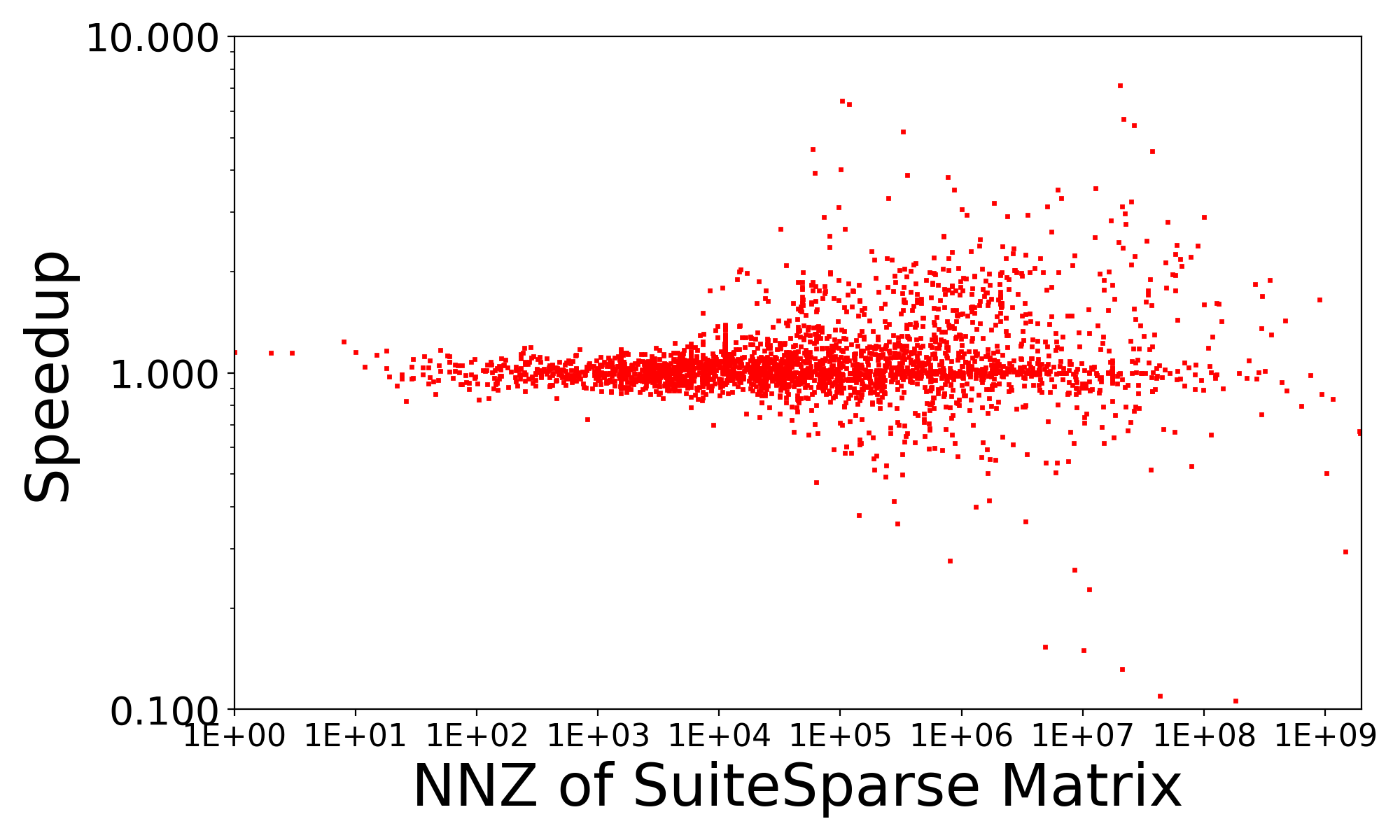}
    \caption{parallel SpMM-lexi}
    \label{fig:spmm_multithreaded_ordering}
\end{subfigure}
\caption{Performance of Lexi ordering}
  \label{fig:lexi_performance}
\end{figure*}
\textbf{SpMV and SpMM.}
We measured the performance of \name and TACO while running SpMV and SpMM with each of the 2833 matrices for sequential and parallel execution. 
We present the experimental results in Figure~\ref{fig:cpu_performance}, where
\name and TACO are represented in red and blue dots respectively. 
In the plot, the x-axis represents a matrix (2,833 matrices, ordered by increasing number of nonzeros) and the y-axis execution time (lower is better). 
As we can see from the plots, \name achieves better performance than TACO on sequential SpMM (Figure~\ref{fig:spmm_seq} and parallel SpMV (Figure~\ref{fig:spmv_multithreaded}, and comparable performance on sequential SpMV and parallel SpMM (Figures~\ref{fig:spmv_seq} and~\ref{fig:spmm_multithreaded}, respectively).
For sequential execution, \name{} outperforms TACO by up to 6.26x for SpMM (average 2.29x) and by up to 2.14x for SpMV (average 0.94x).
A comparison of \name and TACO generated LLVM IR codes shows that \name results in more optimized code with better SIMD (or vectorization) utilization 
 and loop unrolling. For both SpMV and SpMM, take SpMM as an example. The utilization of many SIMD instructions in TACO is only half of that in \name (e.g., TACO only uses 2 lanes while \name uses 4 lanes). 
 \name{} unrolls multiple loops by 8 while TACO unrolls them by 2. 
Although the generated LLVM \gls{IR} for both SpMV and SpMM show similar differences, the effect of better vectorization and loop unrolling are more evident for larger computation (SpMM).
These results highlight one of the major goals of \gls{MLIR} and \gls{MLIR}-based compilers: by leveraging higher-level semantics information and progressive lowering steps, it is possible to produce a more aggressive and higher-quality LLVM \gls{IR} that, eventually, results in higher performance and resource utilization.

For parallel SpMV, \name{} achieves an average of 20.92x speedup over TACO. Especially for small matrices, \name{} outperforms TACO by a significant margin, however, after further inspection, we realized that this performance difference is due to the overhead introduced by the underlying parallel runtime. \name{} uses an asynchronous task-based programming model based on LLVM co-routines while TACO leverages OpenMP. For small computation, LLVM co-routines introduce less overhead than OpenMP threading (which is beneficial for larger parallel regions). As we can see from Figure~\ref{fig:spmm_multithreaded}, when there is enough computation for each OpenMP thread, the runtime overhead is amortized and both \name{} and TACO perform similarly.

\begin{figure}[htbp]
    \includegraphics[width=0.45\textwidth]{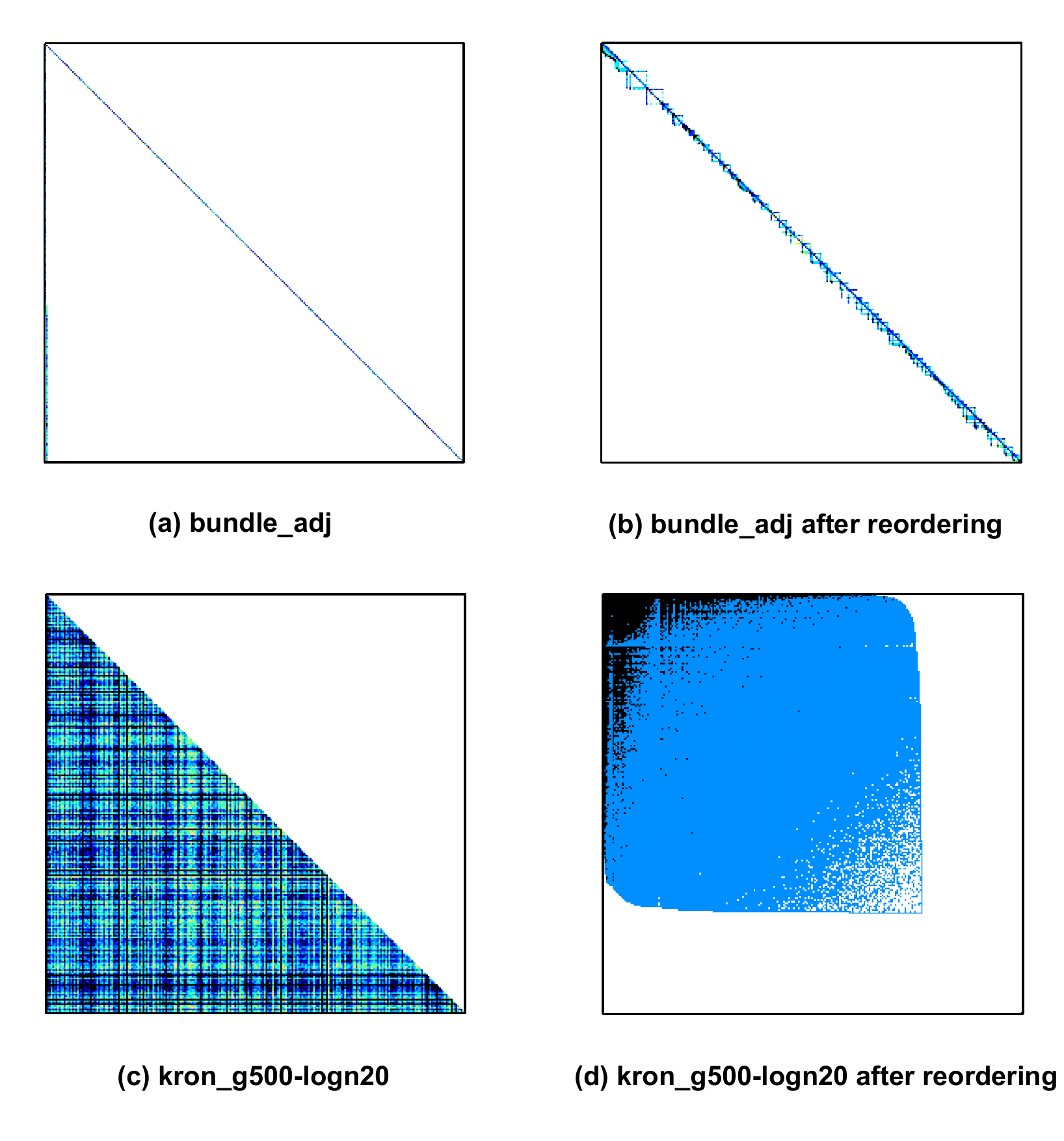}
    \caption{Visualization comparison of matrices with and without reordering}
    \label{fig:vis_reordering}
\end{figure}

\begin{figure*}[htbp]
\begin{subfigure}[b]{0.247\textwidth}
    \includegraphics[width=\textwidth]{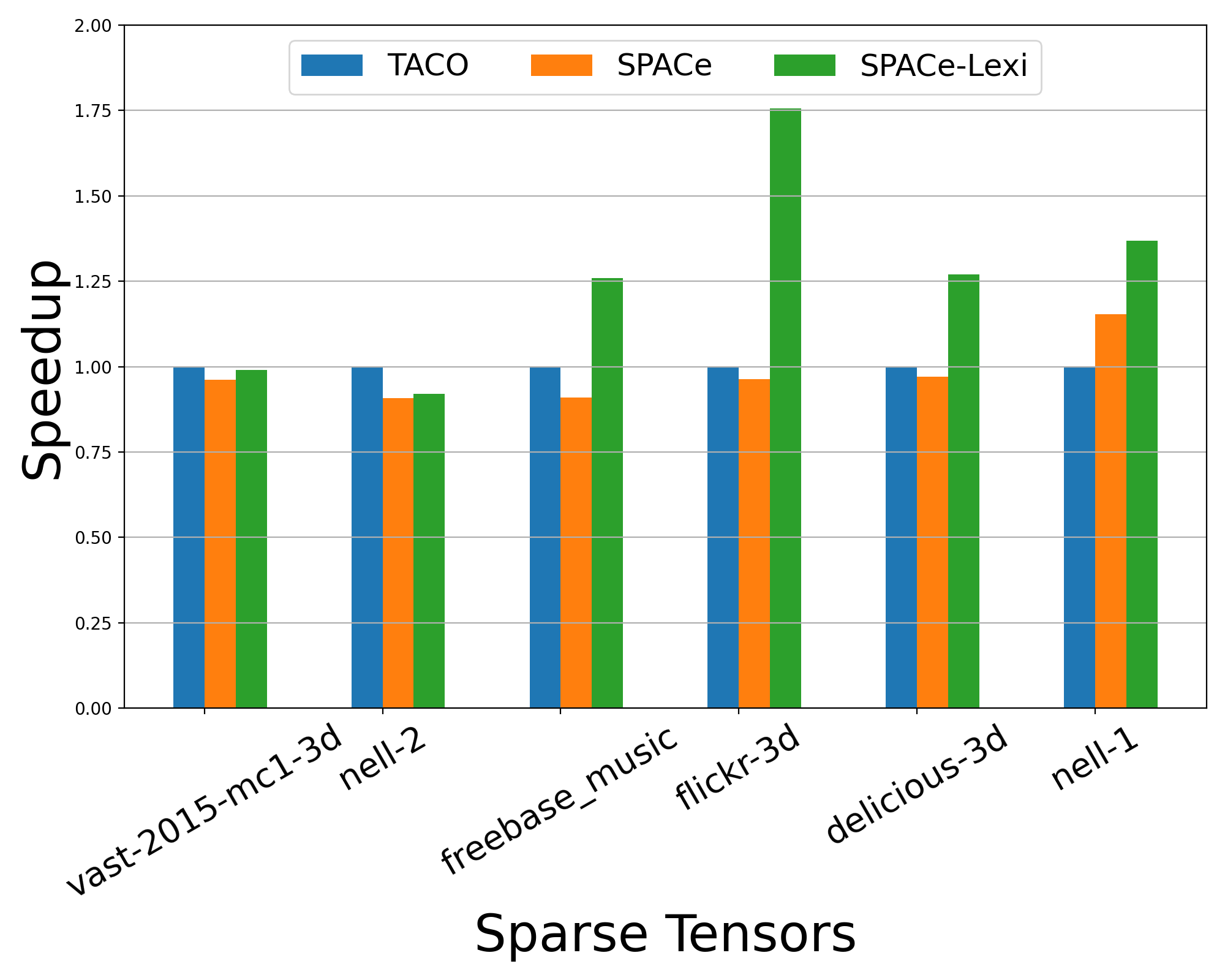}
    \caption{sequential TTV}
    \label{fig:spmv_seq_ordering}
\end{subfigure}
\begin{subfigure}[b]{0.247\textwidth}
    \includegraphics[width=\textwidth]{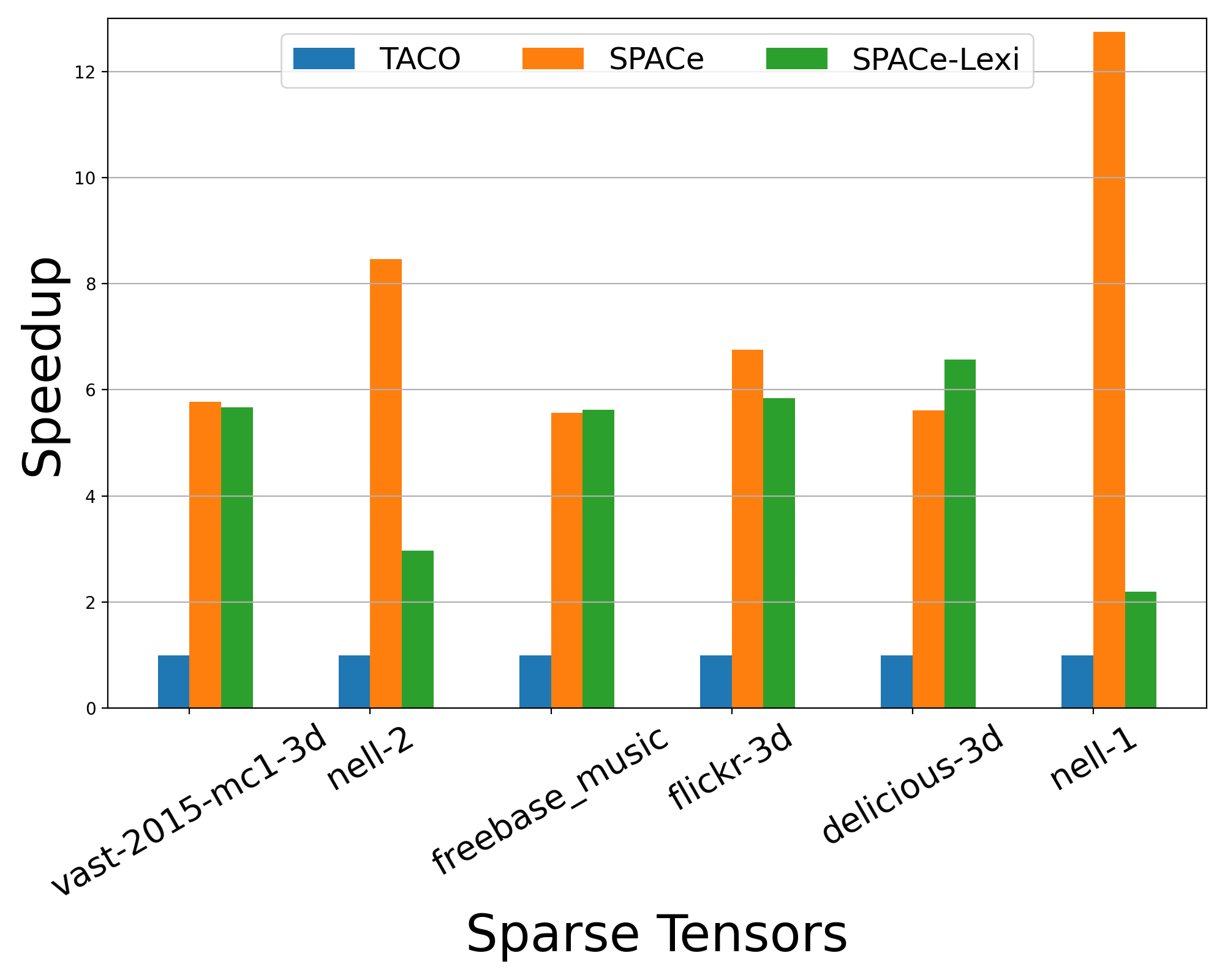}
    \caption{parallel TTV}
    \label{fig:spmv_multithreaded_ordering}
\end{subfigure}
\begin{subfigure}[b]{0.247\textwidth}
    \includegraphics[width=\textwidth]{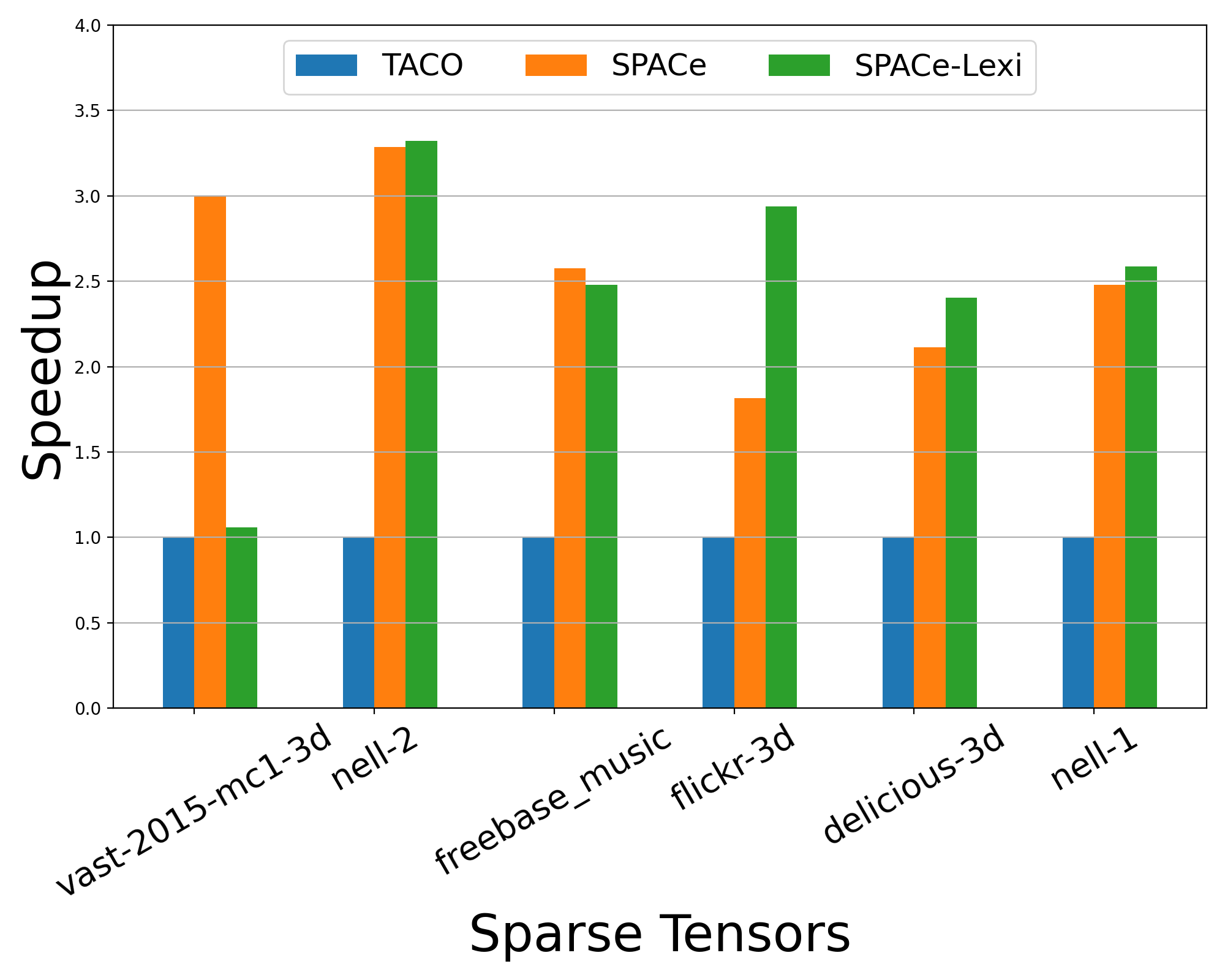}
    \caption{sequential TTM}
    \label{fig:spmv_seq_ordering}
\end{subfigure}
\begin{subfigure}[b]{0.247\textwidth}
    \includegraphics[width=\textwidth]{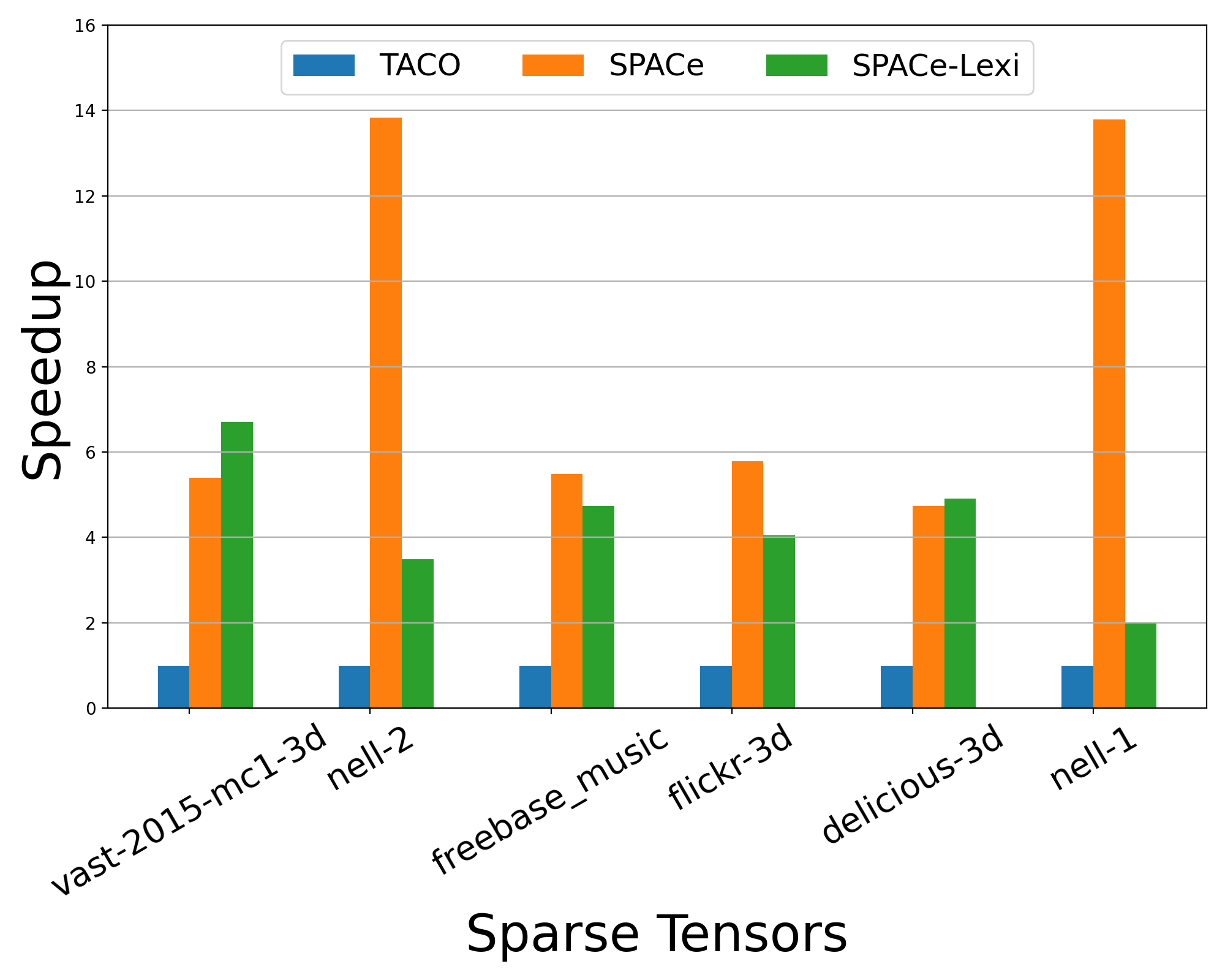}
    \caption{parallel TTM}
    \label{fig:spmv_multithreaded_ordering}
\end{subfigure}
\caption{Performance of tensor operations}
  \label{fig:ttvm_performance}
\end{figure*}

\textbf{Reordering.} By reordering data in memory, \name{} attempts to increase spatial and temporal locality to achieve higher performance.  The plots in Figure~\ref{fig:lexi_performance} show \name{} performance when reordering data compared to original case (no reordering). Figure~\ref{fig:lexi_performance} shows that, indeed, in many cases there is significant advantage of reordering data, with up to 3.41x (average 1.04x), 3.89x (average 1.03x), 7.12x (average 1.12x), and 7.14x (average 1.13x) for SpMV sequential, SpMV parallel, SpMM sequential, and SpMM parallel, respectively.
However, we also note that there might be significant performance degradation, especially for parallel execution. We further analyzed the reasons for this disparity and identified load imbalance as the primary source of performance degradation. Our reordering algorithm attempts to cluster nonzeros on the top-left corner of sparse matrices. In an ideal case, after reordering the nonzeros are distributed around the matrix diagonal.

Figure~\ref{fig:vis_reordering} shows a case in which reordering results in high performance improvements. In this case, the nonzero elements originally around the first column are distributed around the diagonal.
Figure~\ref{fig:vis_reordering}, instead, shows a case in which reordering reduces performance. In this case, the nonzeros are clustered around the top-left corner, thus threads that operate on the top rows have more work to perform compared to threads that operate on the bottom rows, which results in load imbalance and performance degradation.

\textbf{TTV and TTM.} 
We also compare \name with TACO on TTV and TTM with six sparse tensors on CPU and multi-threads and with reordering optimization on and off.
Figure~\ref{fig:ttvm_performance} illustrates the experimental results. 
TACO does not generate parallel code if the output tensor is stored in sparse format, even if instructed to do so, thus the results in the Figure for parallel execution are with respect to sequential execution of the TACO benchmarks.
For sequential TTV, \name performs comparably to TACO. 
With reordering, \name achieves better performance on four out of six sparse tensors. 
For parallel TTV, \name performs significantly better than TACO with up to $12.5\times$ and on average $8\times$ speedup. 
With reordering, \name's performance is degraded on five of six sparse tensors except for delicious-3d. 
As for the case of SpMV and SpMM, we observed similar load imbalance issues.
For sequential TTM, \name performs better than TACO with up to $3.3\times$ and on average $2.53\times$ speedup. 
With reordering, \name achieves better performance on three out of six sparse tensors. 
For parallel TTM, \name performs significantly better than TACO with up to $13.9\times$ and on average $8.13\times$ speedup. 
With reordering, \name's performance is degraded on five of six sparse tensors except for vast-2015-mc1-3d.

Our results show that reordering tensors have a significant (positive or negative) impact on performance, more than for matrices. One possible reason is that the \texttt{LexiOrder} algorithm reorders all dimensions of data simultaneously, which means the data locality is the best when accessing all the dimensions in conjunction, as in conjunction.
The sparse tensor operation MTTKRP~\cite{li2019efficient} follows this behavior to gain a good performance speedup. 
However, this does not mean that the indices in every dimension get good locality when accessing the vector or matrix in TTV or TTM, potentially leading to low performance.
We will investigate alternative reordering algorithms and adaptive methods in future work.

%% file: ops-formats.tex
%% file: related_work.tex
\section{Related Work}
\label{sec:relatedwork}


\textbf{Compiler for Tensor Algebra.}
Compiler techniques have been used to drive irregular computation in tensor algebra~\cite{hirata2003tensor,kjolstad2017tensor,baghdadi2019tiramisu,kim2019code,springer2016ttc}.
TCE~\cite{hirata2003tensor} is a compiler optimization framework that focuses on dense tensor contraction operations in quantum chemistry. 
TTC~\cite{springer2016ttc} is a compiler framework that carries out a composition of high-performance tensor transpose strategies for GPUs.
TACO~\cite{kjolstad2017tensor} is a compiler that generates code for given tensor algebra expressions and used as a higher-level domain-specific language for tensor algebra. 
Kim et al.~\cite{kim2019code} use similar compiler techniques for high-performance tensor contractions but focus on its application on \gls{GPU}s. 
Different from existing works, we develop a high-performance sparse tensor algebra compiler using MLIR, which supports both serial and parallel code generation and enables better portability and adaptability.

\textbf{Domain-specific Libraries for Tensor Algebra.}
There have been a collection of tensor algebra libraries developed~\cite{gunnels2001flame,poulson2013elemental,wang2014intel,schatz2016parallel,schatz2014exploiting,epifanovsky2013new,fishman2020itensor,roberts2019tensornetwork,solomonik2014massively}. 
FLAME~\cite{gunnels2001flame} is a library aiming for the derivation and implementation of tensor algebra operations on CPUs. 
Later, serial linear algebra libraries are extended to run on distributed parallel systems~\cite{poulson2013elemental,schatz2016parallel,schatz2014exploiting,epifanovsky2013new}. 
On the other hand, these libraries are extended to support sparse tensor algebra operations using different sparse tensor formats~\cite{fishman2020itensor,roberts2019tensornetwork,solomonik2014massively}. 
Tensor algebra libraries favor scientific computing and are widely utilized in scientific application development. 
By contrast, \name transparently implements tensor algebra algorithms per se and can compile most types of sparse tensor formats and automatically generate efficient code. 

\textbf{Tensor Algebra Optimization.}
Plenty of work~\cite{auer2006automatic,bell2009implementing,baskaran2014low,kourtis2011csx,smith2015splatt,yang2018design,li2018hicoo, bik1993} leverage reordering to optimize tensor algebra with respect to distinct tensor formats for different tensor operations and heterogeneous architectures. 
Kjolstad et al.~\cite{kjolstad2019tensor,kjolstad:2017:taco} reorder loops of tensor algebra computations to improve the data locality.
Smith et al.~\cite{smith2015splatt} use reordering to enable high-performance tensor factorization operations.
Yang et al.~\cite{yang2018design} identify an efficient memory access pattern for high-performance SpMM operations through merge-based load balancing and row-major coalesced memory access.  
Other works, such as \cite{maggioni2013adell,li2018hicoo,choi2010model,bell2008efficient}, to name a few, design high-performance algorithms considering computer architecture characteristics using techniques like register blocking, cache blocking, and reordering.
\name. 



%% file: conclusion.tex
\section{Conclusion}
\label{sec:conclusion}

In this work, we present a high-performance sparse tensor algebra compiler, called \name, and a high-productive DSL to support next-generation tensor operations.
Our DSL enables high-level programming abstractions that resemble the familiar Einstein notation to express tensor algebra operations. 
\name is based on the MLIR framework, which allows us to build portable, adaptable, and extensible compilers. 
\name provides an effective and efficient code generation which supports most tensor storage formats through an internal storage format based on four dimension attributes and a novel code generation algorithm. 
Furthermore, we incorporate a data reordering algorithm to increase the data locality.
The evaluation results reveal that \name outperforms competing for baseline sparse tensor algebra compiler TACO with up to 20.92x, 6.39x, and 13.9x performance improvement for SpMV, SpMM, and TTM computations respectively. 
In future work, we plan to extend \name{} to support heterogeneous architectures and to explore alternatives reordering schemes that better adapt to the sparsity patterns observed in scientific and engineering input sets.
%

